 \documentclass[longauth]{aa}
\usepackage{graphicx}
\usepackage{natbib}
\usepackage[colorlinks=false]{hyperref}

\def\aap{A\&A}

\def\apjl{ApJL}

\def\vinfty{v_{\infty}\,}
\def\mdot{$\dot{\rm M}$}
\def\kms{km~s$^{-1}$}

\def\rsun{R$_{\odot}$}
\def\msunyr{M$_{\odot}$\,yr$^{-1}$}

\def\LUAN{1}
\def\LAOG{2}
\def\OCA{3}
\def\OAA{4}
\def\MPIFR{5}
\def\StewObs{6}
\def\CRAL{7}
\def\ESOChile{8}
\def\ESOGarching{9}
\def\IRCOM{10}
\def\INSU{11}
\def\ONERA{12}
\def\KULeuven{13}
\def\ObsToulouse{14}
\def\CALERN{15}

\begin{document}
\title{
  Direct constraint on the distance of $\gamma^2$~Velorum from
  AMBER/VLTI observations.
  \thanks{Based on observations made with the Very Large Telescope
    Interferometer at Paranal Observatory}
}

\titlerunning{$\gamma^2$~Velorum observed with AMBER/VLTI}

\institute{
  Laboratoire Universitaire d'Astrophysique de Nice, UMR 6525
  Universit\'e de Nice/CNRS, Parc Valrose, F-06108 Nice cedex 2,
  France
  \and
  Laboratoire d'Astrophysique de Grenoble, U.M.R. 5571 Universit\'e
  Joseph Fourier/C.N.R.S., BP 53, F-38041 Grenoble Cedex 9, France
  \and
  Laboratoire Gemini, U.M.R. 6203 Observatoire de la C\^ote
  d'Azur/C.N.R.S., Avenue Copernic, 06130 Grasse, France
  \and
  INAF-Osservatorio Astrofisico di Arcetri, Istituto Nazionale di
  Astrofisica, Largo E.~Fermi 5, I-50125 Firenze, Italy
  \and
  Max-Planck-Institut f\"ur Radioastronomie, Auf dem H\"ugel 69, D-53121
  Bonn, Germany
  \and
  Steward Observatory, University of Arizona, 933 North Cherry Avenue,
  Tucson, AZ 85721, USA 
  \and
  Centre de Recherche Astronomique de Lyon, U.M.R. 5574 Universit\'e
  Claude Bernard/C.N.R.S., 9 avenue Charles Andr\'e, F-69561 Saint Genis
  Laval cedex, France
  \and
  European Southern Observatory, Casilla 19001, Santiago 19, Chile
  \and
  European Southern Observatory, Karl Schwarzschild Strasse 2, D-85748
  Garching, Germany
  \and
  IRCOM, U.M.R. 6615 Universite de Limoges/C.N.R.S., 123 avenue Albert
  Thomas, F-87060 Limoges cedex, France
  \and
  Division Technique INSU/C.N.R.S. UPS 855, 1 place Aristide Briand,
  F-92195 Meudon cedex, France
  \and
  ONERA/DOTA, 29 av de la Division Leclerc, BP 72, F-92322 Ch\^atillon
  Cedex, France
  \and
  Instituut voor Sterrenkunde, KULeuven, Celestijnenlaan 200B, B-3001
  Leuven, Belgium
  \and
  \emph{Present affiliation:} Laboratoire Astrophysique de
  Toulouse, UMR 5572 Universit\'e Paul Sabatier/CNRS, BP 826, F-65008
  Tarbes cedex, France 
  \and
  \emph{Present affiliation:} Observatoire de la C�e d'Azur -
  Calern, 2130 Route de l'Observatoire , F-06460 Caussols, France
}

\author{
  F.~Millour\inst{\LUAN;\LAOG}
  \and R.~G.~Petrov\inst{\LUAN}
  \and O.~Chesneau\inst{\OCA}
  \and D.~Bonneau\inst{\OCA}
  \and L.~Dessart\inst{\StewObs}
  \and C.~Bechet\inst{\CRAL}
  \and I.~Tallon-Bosc\inst{\CRAL}
  \and M.~Tallon\inst{\CRAL}
  \and E.~Thi\'ebaut\inst{\CRAL}
  \and F.~Vakili\inst{\LUAN}
  \and F.~Malbet\inst{\LAOG}
  \and D.~Mourard\inst{\OCA}
  \and 
\\G.~Zins\inst{\LAOG}
  \and A.~Roussel\inst{\OCA}
  \and S.~Robbe-Dubois\inst{\LUAN}
  \and P.~Puget\inst{\LAOG}
  \and K.~Perraut\inst{\LAOG}
  \and F.~Lisi\inst{\OAA}
  \and E.~Le~Coarer\inst{\LAOG}
  \and S.~Lagarde\inst{\OCA}
  \and P.~Kern\inst{\LAOG}
  \and L.~Gl\"uck\inst{\LAOG}
  \and G.~Duvert\inst{\LAOG}
  \and A.~Chelli\inst{\LAOG}
  \and Y.~Bresson\inst{\OCA}
  \and U.~Beckmann\inst{\MPIFR}
  \and P.~Antonelli\inst{\OCA}
  \and 
\\G.~Weigelt\inst{\MPIFR}
  \and N.~Ventura\inst{\LAOG}
  \and M.~Vannier\inst{\LUAN;\ESOChile}
  \and J.-C.~Valtier\inst{\OCA}
  \and L.~Testi\inst{\OAA}
  \and E.~Tatulli\inst{\LAOG;\OAA}
  \and D.~Tasso\inst{\OCA}
  \and P.~Stefanini\inst{\OAA}
  \and P.~Stee\inst{\OCA}
  \and W.~Solscheid\inst{\MPIFR}
  \and D.~Schertl\inst{\MPIFR}
  \and P.~Salinari\inst{\OAA}
  \and M.~Sacchettini\inst{\LAOG}
  \and A.~Richichi\inst{\ESOGarching}
  \and F.~Reynaud\inst{\IRCOM}
  \and S.~Rebattu\inst{\OCA}
  \and Y.~Rabbia\inst{\OCA}
  \and T.~Preibisch\inst{\MPIFR}
  \and C.~Perrier\inst{\LAOG}
  \and F.~Pacini\inst{\OAA}
  \and K.~Ohnaka\inst{\MPIFR}
  \and E.~Nussbaum\inst{\MPIFR}
  \and D.~Mouillet\inst{\LAOG,\ObsToulouse}
  \and J.-L.~Monin\inst{\LAOG}
  \and P.~M\`ege\inst{\LAOG}
  \and P.~Mathias\inst{\OCA}
  \and G.~Martinot-Lagarde\inst{\INSU,\CALERN}
  \and G.~Mars\inst{\OCA}
  \and A.~Marconi\inst{\OAA}
  \and Y.~Magnard\inst{\LAOG}
  \and B.~Lopez\inst{\OCA}
  \and D.~Le~Contel\inst{\OCA}
  \and J.-M.~Le~Contel\inst{\OCA}
  \and S.~Kraus\inst{\MPIFR}
  \and D.~Kamm\inst{\OCA}
  \and K.-H.~Hofmann\inst{\MPIFR}
  \and O.~Hernandez~Utrera\inst{\LAOG}
  \and M.~Heininger\inst{\MPIFR}
  \and M.~Heiden\inst{\MPIFR}
  \and C.~Gil\inst{\LAOG}
  \and E.~Giani\inst{\OAA}
  \and A.~Glentzlin\inst{\OCA}
  \and S.~Gennari\inst{\OAA}
  \and A.~Gallardo\inst{\LAOG}
  \and D.~Fraix-Burnet\inst{\LAOG}
  \and R.~Foy\inst{\CRAL}
  \and E.~Fossat\inst{\LUAN}
  \and T.~Forveille\inst{\LAOG}
  \and D.~Ferruzzi\inst{\OAA}
  \and P.~Feautrier\inst{\LAOG}
  \and M.~Dugu\'e\inst{\OCA}
  \and T.~Driebe\inst{\MPIFR}
  \and A.~Domiciano~de~Souza\inst{\LUAN}
  \and A.~Delboulb\'e\inst{\LAOG}
  \and C.~Connot\inst{\MPIFR}
  \and J.~Colin\inst{\OCA}
  \and J.-M.~Clausse\inst{\OCA}
  \and F.~Cassaing\inst{\ONERA}
  \and S.~Busoni\inst{\OAA}
  \and S.~Bonhomme\inst{\OCA}
  \and T.~Bl\"ocker\inst{\MPIFR}
  \and J.~Behrend\inst{\MPIFR}
  \and C.~Baffa\inst{\OAA}
  \and E.~Aristidi\inst{\LUAN}
  \and B.~Arezki\inst{\LAOG}
  \and K.~Agabi\inst{\LUAN}
  \and B.~Acke\inst{\LAOG;\KULeuven}
  \and M.~Accardo\inst{\OAA}
  \and M.~Kiekebusch\inst{\ESOChile}
  \and F.~Rantakyr\"o\inst{\ESOChile}
  \and M.~Sch\"oller\inst{\ESOChile}
}

\offprints{
  F.~Millour\\
  \email{Florentin.Millour@unice.fr}
}

\date{Received; accepted}

\abstract
{
  {Interferometry can provide spatially {\it resolved} observations of
    massive star binary systems and their colliding winds, which thus far
    have been studied mostly by means of spatially {\it unresolved}
    observations.}
}
{
  In this work, we present the first AMBER/VLTI observations, taken at
  orbital phase 0.32, of the Wolf-Rayet and O (WR+O) star binary
  system $\gamma^2$~Velorum and use 
  the interferometric observables to constrain its properties.
}
{
  The AMBER/VLTI instrument was used with the telescopes UT2,
    UT3, and UT4 on baselines ranging from 46\,m to 85\,m. It
    delivered spectrally dispersed visibilities, as well as
    differential and closure phases, with a resolution $R=1500$ in the
    spectral band 1.95-2.17\,$\mu$m. We interpret these data in the
    context of a binary system with unresolved components, neglecting
    in a first approximation the wind-wind collision zone flux
    contribution.
}
{
  Using WR- and O-star synthetic spectra, we show that the AMBER/VLTI
  observables result primarily from the contribution of the individual
  components of the WR+O binary system. We discuss several
  interpretations of the residuals, and speculate on the detection of
  an additional 
  continuum component, originating
  from the free-free emission associated with the wind-wind collision
  zone (WWCZ), and contributing at most to the observed $K$-band flux at the
  5\% level. Based on the accurate spectroscopic orbit and the
  Hipparcos distance, the expected absolute separation and position
  angle at the time of observations were 5.1$\pm$0.9mas and
  66$\pm$15\degr, respectively
  . However, using theoretical estimates
  for the spatial extent of both continuum and  line emission from
  each component, we infer a separation of 3.62$^{+0.11}_{-0.30}$\,mas
  and a position angle of 73$^{+9}_{-11}$\degr, compatible with
    the expected one.  Our analysis thus
  implies that the binary system lies at a distance of
  368$^{+38}_{-13}$\,pc, in agreement with recent
  spectrophotometric estimates, but significantly larger than the
  Hipparcos value of 258$^{+41}_{-31}$\,pc.
}
{}

\keywords{Techniques: interferometric --
  stars: individual: $\gamma^2$~Velorum - stars: winds, outflows -
  stars: Wolf-Rayet
  stars: binaries: spectroscopic - stars: early-type }

\maketitle

%

\section{Introduction}
\label{section:intro}

$\gamma^2$~Velorum constitutes an excellent laboratory for the study of 
massive stars and their radiatively-driven winds.
Indeed, $\gamma^2$~Velorum (WR 11, HD 68273) is the closest known Wolf-Rayet
(WR) star, at a Hipparcos-determined distance of 258$^{+41}_{-31}$\,pc
\citep{1997NewA....2..245V,   1997ApJ...484L.153S}, whereas other WR objects 
lie at $\sim$1\,kpc or beyond. 
Moreover, $\gamma^2$~Velorum is an SB2 spectroscopic binary WR+O system 
\citep[WC8+O7.5{\sc iii}, P = 78.53 d,][]{1997A&A...328..219S,
  1999A&A...345..163D} offering access to fundamental parameters
of the WR star, usually obtained indirectly through
the study of its dense and fast wind. Using spectroscopic modeling of
the integrated, but spectrally-dispersed, light from the system,
\citet{1999A&A...345..163D} and \citet{2000A&A...358..187D} provided
the most up-to-date fundamental parameters of the individual
components of the binary system.  

Since $\gamma^2$~Velorum
is relatively bright and well observable at any wavelength, it has
been extensively studied by various techniques. 
In that context, $\gamma^2$~Velorum represents
an unique opportunity to spatially resolve a WR wind by means of
optical interferometry. This object was observed by the
Narrabri intensity interferometer operating around 0.45\,$\mu$m as early
as 1968 \citep{1970MNRAS.148..103H}. By observing such a star with a
long baseline interferometer, one may constrain various parameters, such as
the binary orbit, the brightness ratio of the two components, the angular size 
associated with both the continuum and the lines emitted by the WR star.

The collision between the fast and dense wind from the WR and the less
dense but faster wind from the O star generates a wealth of phenomena.
International Ultraviolet Explorer (IUE) and Copernicus
ultraviolet spectra \citep[and references therein]{1993ApJ...415..298S}
revealed a variability in UV P-Cygni line profiles, associated
with selective line eclipses of the O star light by the WR star wind
as well as the carving of the WR wind by the O-star wind (compared to
its spherical distribution in the absence of a companion). Air-borne
X-ray observation campaigns have revealed additional and invaluable
information on  the wind-wind collision zone (hereafter WWCZ)
\citep[see references in][]{2002Ap&SS.281..199V, 1995A&A...298..549W,
  2001ApJ...558L.113S, 2002A&A...388L..20P, 2003AAS...203.5802C,
  2004A&A...422..177S, 2005MNRAS.356.1308H}.


The WR component of the $\gamma^2$~Velorum system is of a WC8
type. While 50\% of such WR stars (and 90\% of the WC9 type) show
heated ($T_d\sim1300K$) circumstellar amorphous carbon dust, ISO
observations of $\gamma^2$~Velorum revealed no such dust signatures
\citep{1997NewA....2..245V}. Keck observations resolved, although only
barely, the system in the $K$ band and confirmed the absence of any
dust emission from this system \citep{2002ASPC..260..331M}, suggesting
that if dust is created near the WWCZ, it is in small amounts
.

The present paper aims at constraining further our knowledge of the
$\gamma^2$~Velorum system, using long-baseline interferometric
observations conducted in near-IR by the VLTI with the
newly commissioned instrument AMBER.
The results discussed in this paper are limited to
observations recorded with a single triplet of baselines in $K$ band
with the AMBER. We concentrate our
efforts in presenting the potential of AMBER observations and their
complementarity with techniques lacking spatial resolution for the
study of massive close binaries. In this context, we perform an
in-depth checking of the consistency of the AMBER/VLTI data recorded,
with the up-to-date knowledge we have on this well studied binary system.

The paper is organized as follows. In Sect.~\ref{section:dataproc},
we describe the AMBER/VLTI observations and the data. 
In Sect.~\ref{section:binaryparameters}, we present the AMBER data and 
constrain the system characteristics. The system's parameters are used to
predict a basic signal inferred from the knowledge of the
spectroscopic orbit, some estimate of the angular diameter of each
component, and results from the spectroscopic modeling of the individual stars.
We also discuss the potential effects stemming from the WWCZ
and their influence on the observed interferometric signal. 
In Sect.~\ref{section:analysis}, we fit the AMBER observations
by concentrating on a few parameters that can be directly constrained, i.e.,
the angular separation of the components, the orientation of the system
projected onto the sky, and the brightness ratio between the two components. 
We then discuss in Sect.~\ref{section:discussion} the adequacy of our modeling
in reproducing the observations, and present our conclusions in 
Sect.~\ref{section:conclusion}.

\section{AMBER Observations and data}
\label{section:dataproc}

\begin{table*}[htbp]
  \centering
  \caption{
    \footnotesize{
      Log of the observations and atmospheric conditions for
      $\gamma^2$~Velorum (top-three rows) and the spectrophotometric
      calibrator star HD 75063 (bottom-three rows), observed on 25/12/2004.
    }
  }
  \label{obsLog}

  \begin{tabular}{ccccccc}
    \hline

    Time & Star & $K$ Mag. & $\Delta\lambda$ &
    Seeing & Coherence Time & Air Mass\\

    \hline \\

    4h18 & $\gamma^2$~Velorum & 2.1 & 1.95 - 2.03\,$\mu$m &
    0.72\,$\arcsec$ & 4.5\,ms & 1.213 \\

    4h35 & $\gamma^2$~Velorum & 2.1 & 2.02 - 2.10\,$\mu$m &
    0.60\,$\arcsec$ & 5.3\,ms & 1.180 \\

    4h49 & $\gamma^2$~Velorum & 2.1 & 2.09 - 2.17\,$\mu$m &
    0.73 \,$\arcsec$ & 4.4\,ms & 1.158 \\

    \hline \\

    5h57 & HD 75063 & 3.6 & 1.95 - 2.03\,$\mu$m &
    0.90\,$\arcsec$ & 3.6\,ms\,& 1.107 \\

    6h09 & HD 75063 & 3.6 & 2.02 - 2.10\,$\mu$m &
    0.72\,$\arcsec$ & 4.0\,ms & 1.097 \\

    6h18 & HD 75063 & 3.6 & 2.09 - 2.17\,$\mu$m &
    0.58\,$\arcsec$ & 5.5\,ms & 1.090 \\
    \hline
  \end{tabular}
\end{table*}

\begin{figure*}[htbp]
  \centering
  \begin{tabular}{cc}
    \includegraphics[width=0.45\textwidth]{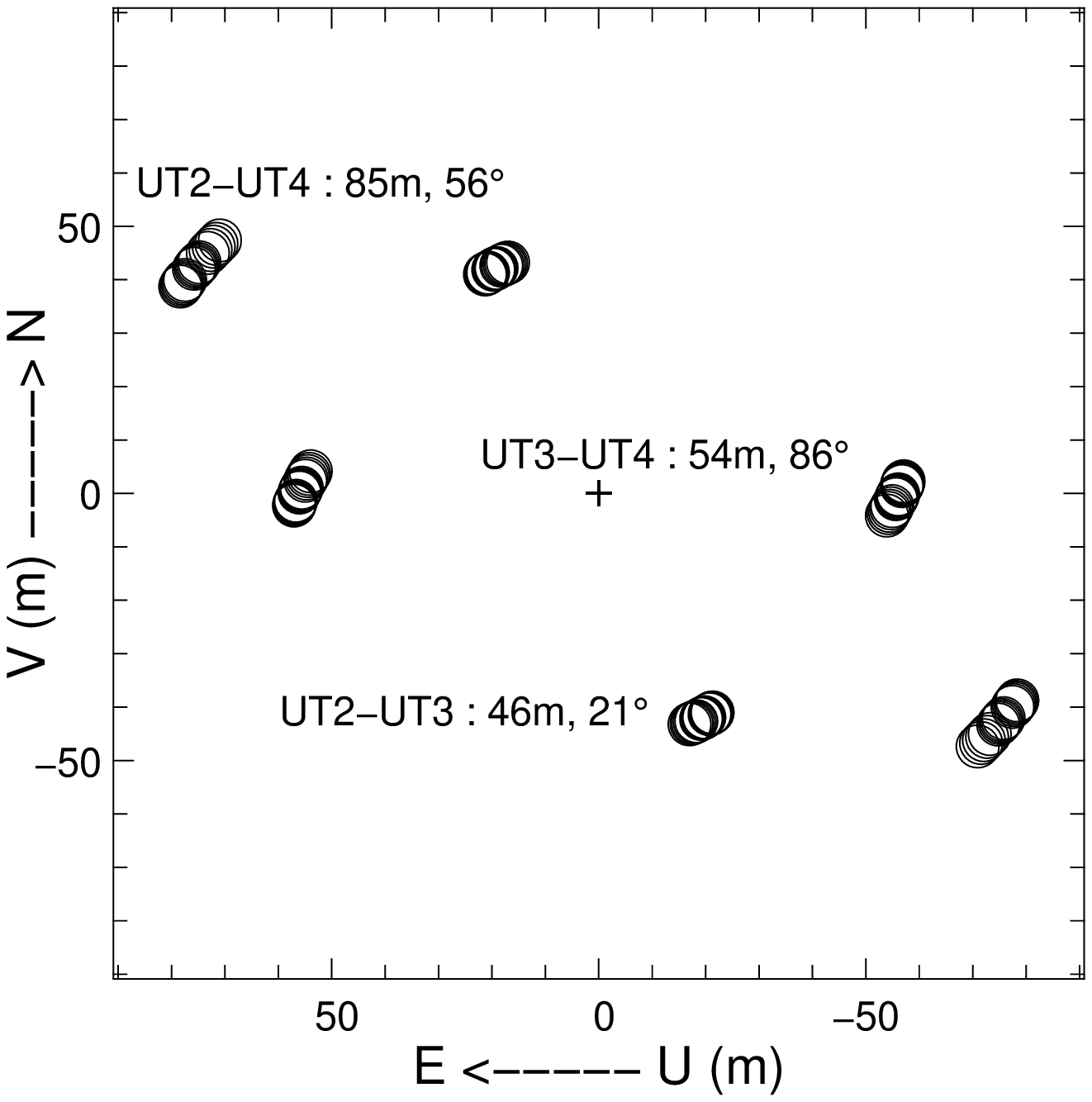}&
    \includegraphics[width=0.45\textwidth, angle=0]{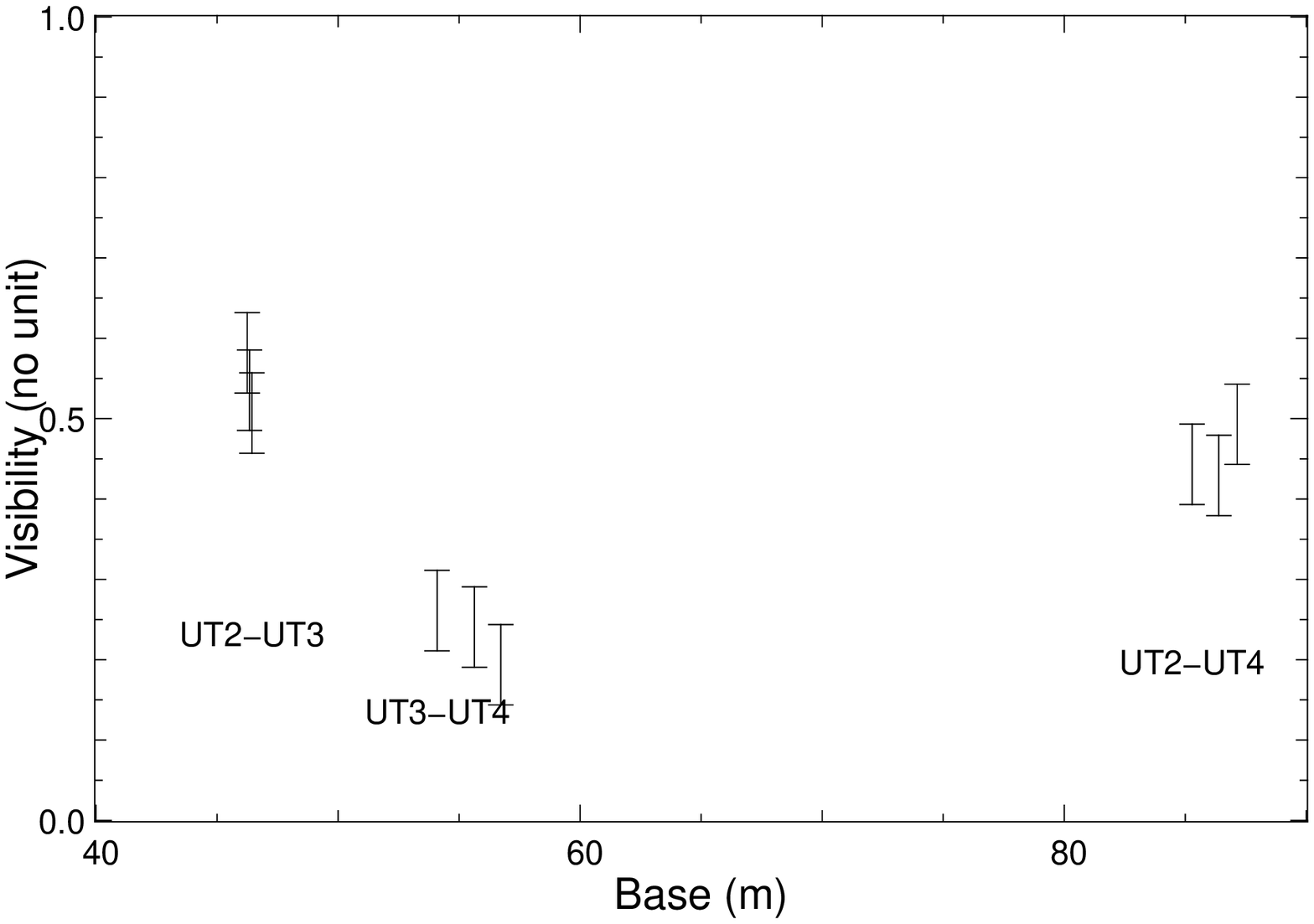}\\
  \end{tabular}
  \caption{
    \footnotesize{
      {\it Left}:
      Projected baselines showing the range of position angles and
      base lengths on the sky during the observations.
      {\it Right}:
      Absolute calibrated visibility of $\gamma^2$~Velorum in function
      of base length. The strong variations correspond mainly to the
      partial resolution of the binary star.
    }
  }
  \label{figure:obs_vis}
\end{figure*}

\subsection{Observations}
\label{subsection:observations}

AMBER (Astronomical Multi BEam Recombiner) is the VLTI (Very Large
Telescope Interferometer) beam combiner operating in the near-infrared
\citep{2003SPIE.4838..924P}. The instrument uses spatial
filtering with fibers \citep{2000SPIE.4006..299M}. The interferometric
beam passes through an anamorphic optics compressing the beam
perpendicularly to the fringe coding in order to be injected into the
slit of a spectrograph. The instrument can operate at spectral
resolutions up to 10,000, and efficiently deliver spectrally dispersed
visibilities.

$\gamma^2$~Velorum was observed on 25 Dec 2004 during the first night
of the first Guaranteed Time Observations (GTO) run of the AMBER
instrument on the three projected baselines UT2-UT3 (46m,\,20\degr),
UT3-UT4 (53m,\,84\degr) and UT2-UT4 (85m,\,55\degr) of the VLTI (see
Fig.~\ref{figure:obs_vis}, left). The Julian day of observation was
JD=2,453,365.16.

The AMBER observations have been conducted with a frame exposure time
of 60\,ms in three spectral windows in the MR-K spectral mode (spectral
resolution of 1500, $K$ band), i.e., 1.98-2.02\,$\mu$m, 2.03-2.11\,$\mu$m, and
2.10-2.17\,$\mu$m at hour angle -134\,min, -114\,min, and -102\,min,
respectively (left panel of Fig.~\ref{figure:obs_vis}).

HD75063 (spectral type A1{\sc iii}) was observed with the same
exposure time and the same spectral windows in order to calibrate the
visibilities. Its diameter is estimated to be 0.50\,mas with an error
of 0.08\,mas, using several color indices. This corresponds to a
visibility of 0.988$\pm$0.004 for the longest base (85m), so that the
error on the calibrator diameter translates into a global error on the
absolute visibility of less than 1\%.

\subsection{Observing context}
\label{subsection:observingContext}

The observations of $\gamma^2$~Velorum were carried out under non
optimal conditions as the VLTI + AMBER system was still not in a
fully operational state at the moment of the observations. As
explained in \citet{2005-AA-MWC297}, a detailed analysis of the
commissioning data has shown that the optical trains of the UT
telescopes are affected by non-stationary high-amplitude
vibrations. These vibrations affect the continuum visibilities,
requiring a careful data processing and calibration procedure. 
We stress that these vibrations bias the instantaneous
estimated visibility but do not affect the closure phase and, since the
observed spectral windows are small, the differential estimators.

During these observations, problems were encountered with the UT2
Adaptive Optics associated with difficulties to close the loop and
with injections in the fibers. We thus expect calibration problems on
the data related to the baselines containing the UT2 telescope. Again,
these problems are limited to the absolute visibilities.

The time lag between the observations of the calibrator and the
science object is of the order of one hour. We checked that the
atmospheric conditions changed only slightly between the two
measurements.



\begin{figure}[htbp]
  \centering
  \begin{tabular}{c}
    \includegraphics[height=0.21\textheight]{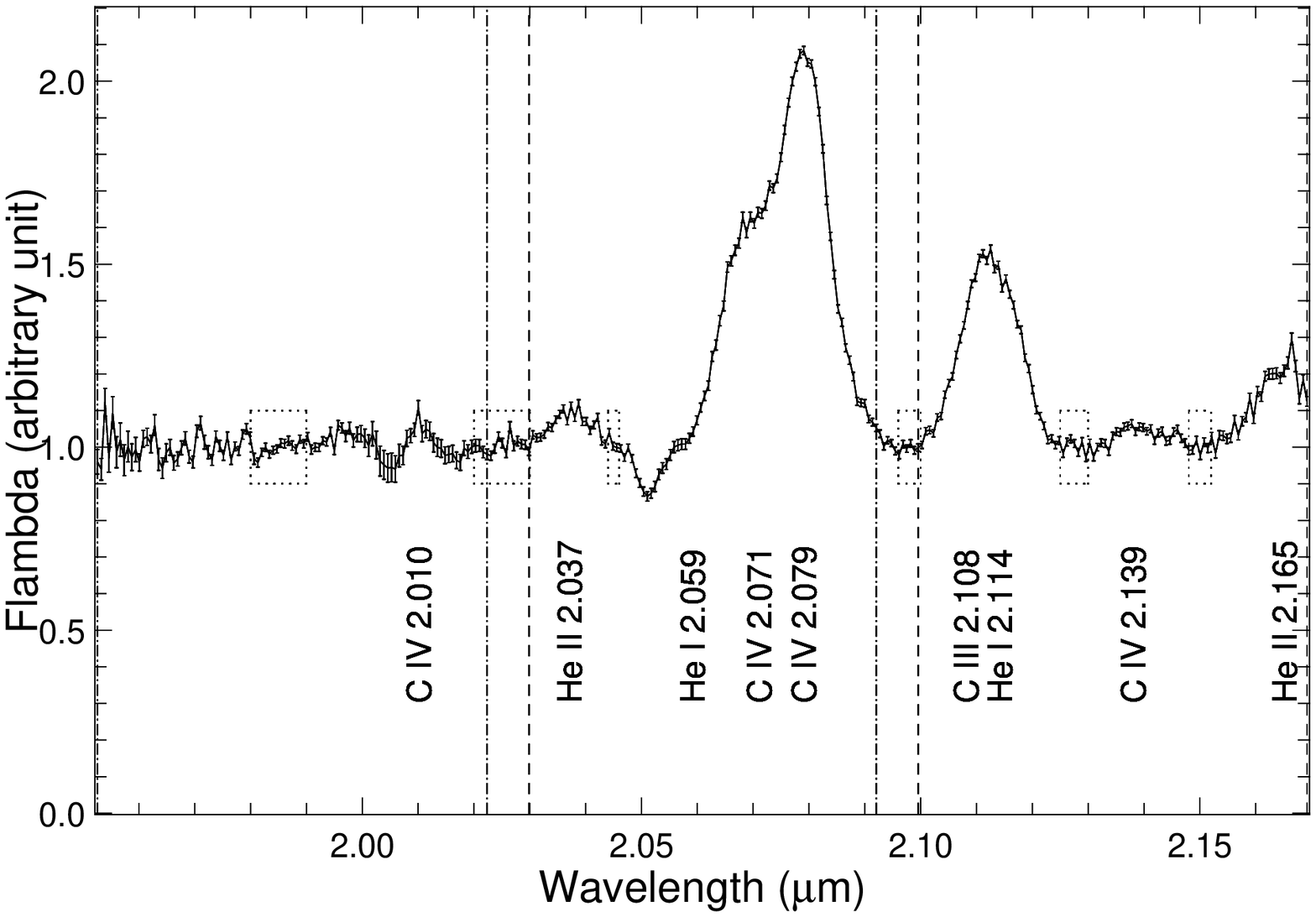}\\
    \includegraphics[height=0.21\textheight]{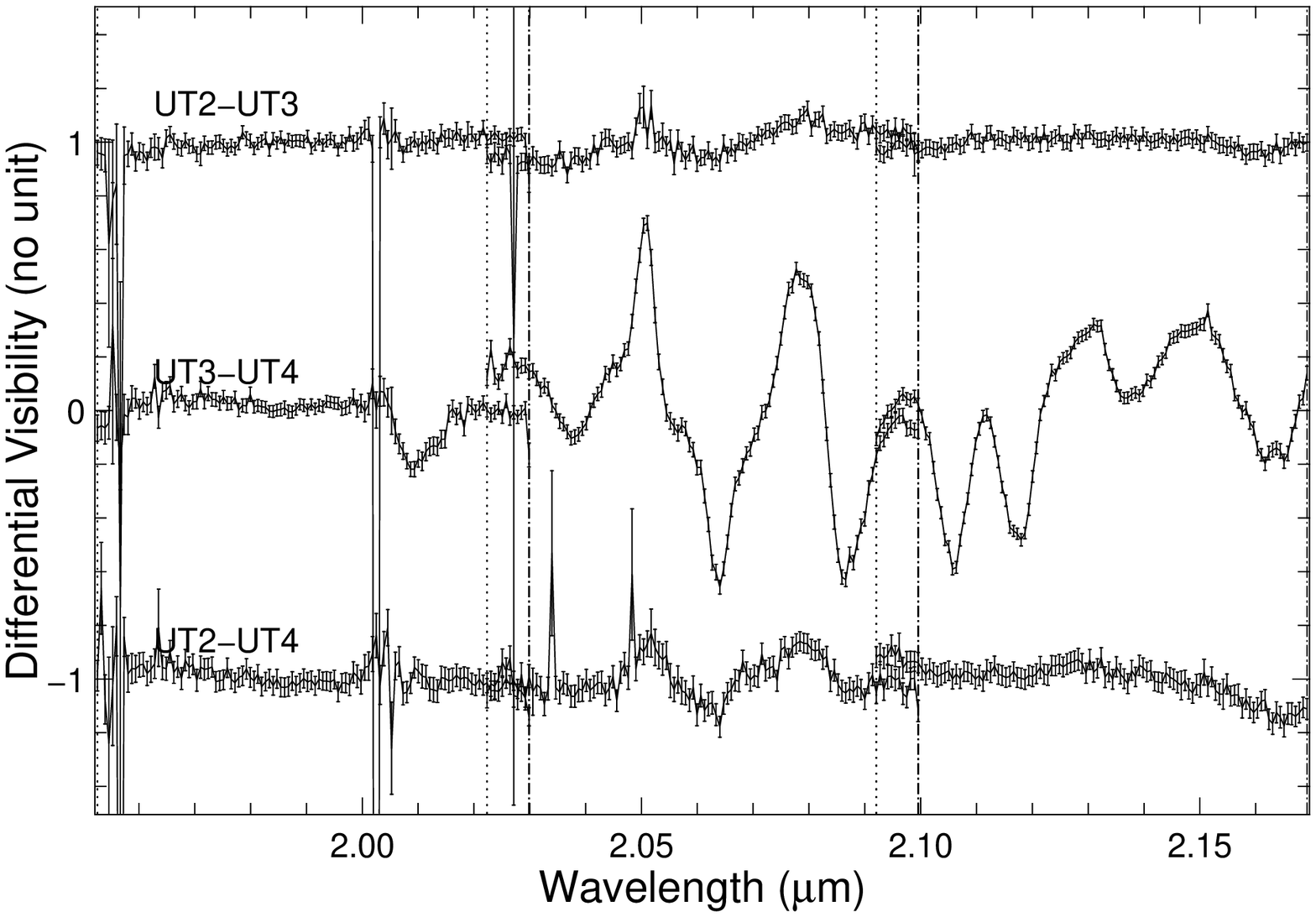}\\
    \includegraphics[height=0.21\textheight]{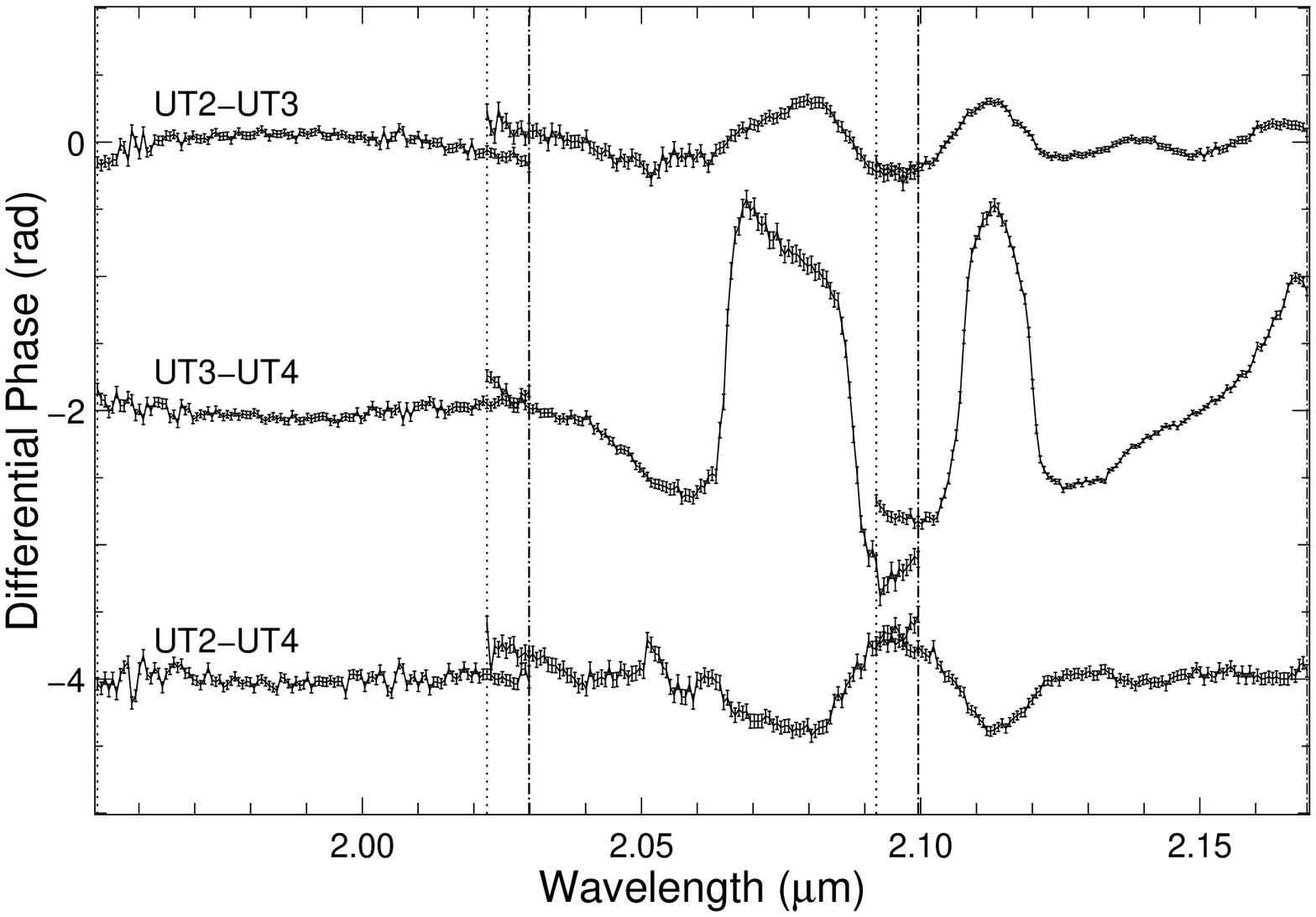}\\
    \includegraphics[height=0.21\textheight]{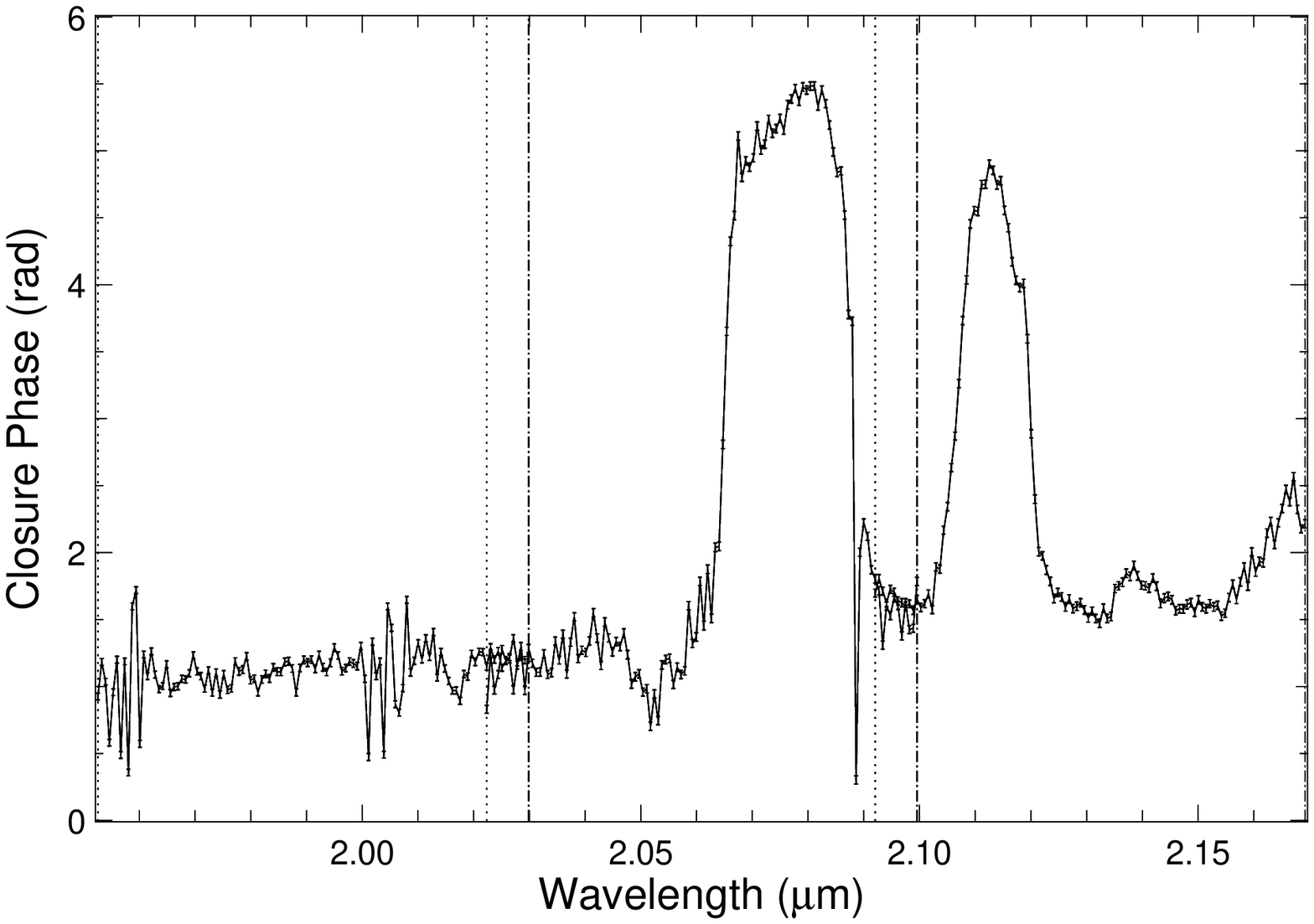}\\
  \end{tabular}
  \caption{
    \footnotesize{
      {\it Top}:
      Result of the complete spectral calibration, showing the
      principal lines present in the observed spectrum. The dotted
      boxes show the selected continuum zones used to redress the
      observed spectrum.
      {\it The other plots from top to bottom}:
      Observed data with AMBER. You can find respectively the
      differential visibilities, the differential phases and the
      closure phase versus wavelength. The differential visibility and
      differential phase curves from each baselines are arbitrarily
      shifted for clarity.
    }
  }
  \label{figure:Spectrum}
\end{figure}


\subsubsection{Data set}
\label{subsubsection:dataSet}

The AMBER/VLTI instrument data processing principles are well
described in the articles of \citet{2006_Tatulli,
  2004SPIE.5491.1222M}. In the current situation, the AMBER/VLTI
instrument has a series of issues about data that obliged us to
develop a specific data reduction strategy which is fully described in
appendices.

The set of data provided by the AMBER instrument (limited to the
spectral window 1.95 to 2.17\,$\mu$m) is the following:

\begin{enumerate}
\item One normalized spectrum (mean spectrum from the three observing
  telescopes).
\item Three absolute visibilities per observation, providing some
  information on the projected equivalent size of the object for each
  baseline.
\item Three differential visibilities curves, providing some
  information on variation of the projected equivalent size of the
  object versus wavelength for each baseline.
\item Three differential phase curves, providing some information on
  the object photocenter relatively to a large continuum reference for
  each baseline.
\item One closure phase providing some information on the degree of
  asymmetry of the system's flux distribution computed for a baseline
  triplet.
\end{enumerate}

The calibrated data are shown in Fig.~\ref{figure:obs_vis} and
Fig.~\ref{figure:Spectrum}.  The differential visibilities,
differential phase and closure phase show strong variations with
wavelength, well beyond the error bars, demonstrating that the
$\gamma^2$~Velorum system has been resolved by the AMBER/VLTI
instrument.

The global slope on the closure phase is mainly due to the
wavelength dependence of spatial frequencies (pure geometrical
effect) whereas the rapid variations in differential visibilities,
differential phases and closure phase are due to variations of the
flux ratio between the two stars (pure spectroscopic effect). On the
contrary, the slope on the differential phases does not have any
physical significance since it depends only on the definition of the
reference channel.

Note that observations in each spectral window were scanned
sequentially, at a quarter-of-an-hour interval. This time lag must
be taken into account since we expect the binary signal to be
rapidly evolving as the triplet of projected baselines slowly
changes due to the earth rotation.

\section{Stars models based on spectroscopic data}
\label{section:binaryparameters}
Below, we present the current knowledge of the system, and use its
geometrical parameters to estimate the basic signal from this binary
system. As a first approach, we neglect any additional emission from
dust; we also neglect the free-free emission expected to arise from
the WWCZ region. This late assumption is probably less valid
physically since the existence of this WWCZ is proved
observationnaly
\footnote{The WWCZ is characterized by a very hot and dense layer of
  material at the interface between the two winds (well above the
  local WR- or O-star wind density) that should contribute
  significantly to the free-free continuum. }.

\begin{figure}[htbp]
  \centering
  \includegraphics[width=0.45\textwidth, angle=0]{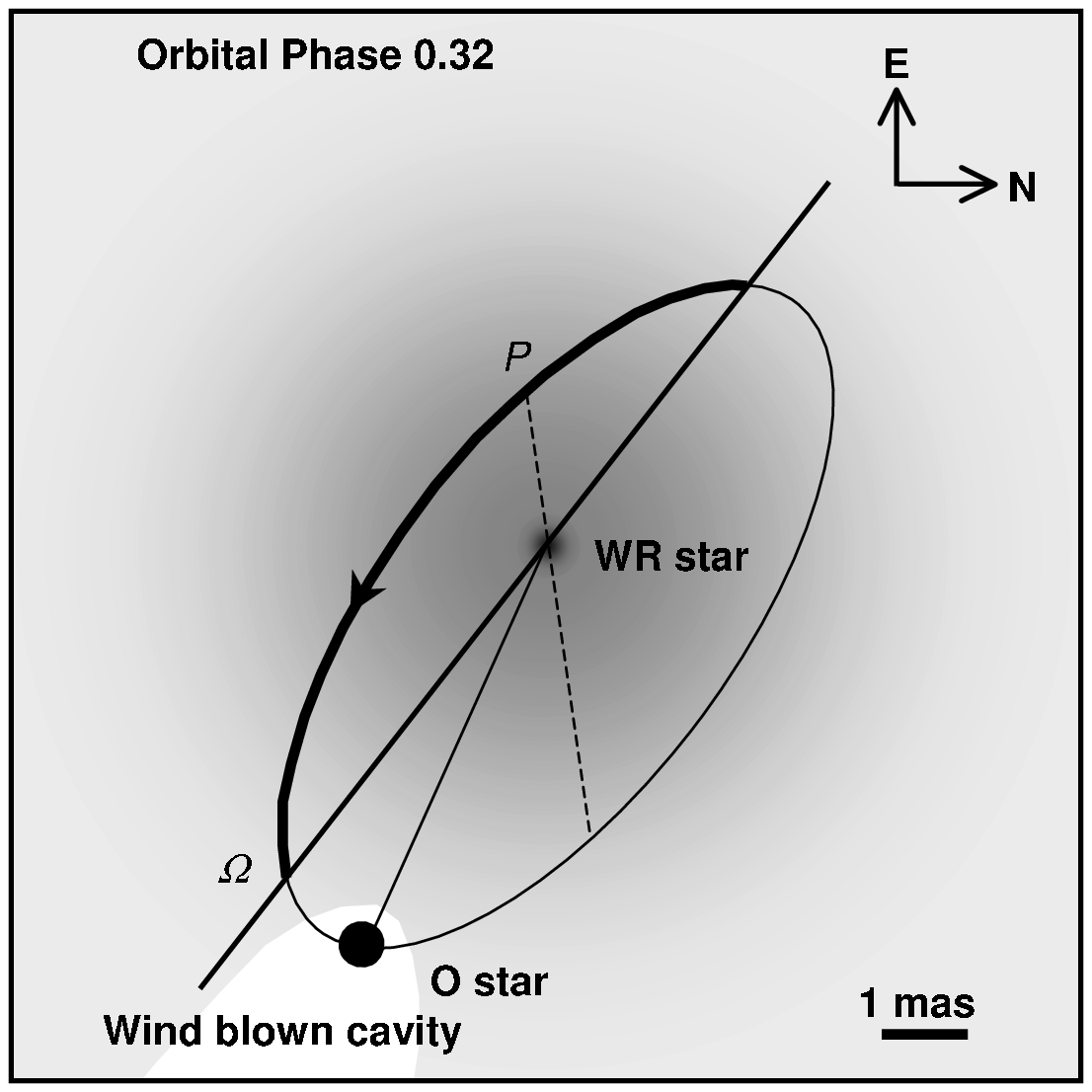}
  \caption{
    Projection of the true orbit as defined by the spectroscopic
    parameters \citep{1997A&A...328..219S} onto the plane of the
    sky. Note that there is an ambiguity of 180\degr on the true
    direction of the WR versus the O star which can be checked over
    by our interferometric observations.
  }
  \label{figure:Orbite}
\end{figure}

\subsection{Geometrical parameters of the system}
\label{subsection:geomParamSys}

The spectroscopic orbit is well determined by
\citet{1997A&A...328..219S}. The interferometer is sensitive to
other geometrical parameters much less constrained, namely the
angular separation and angle of position in the sky.

The observations of \citet{1970MNRAS.148..103H} provided an angular
semi-major axis of the orbit of 4.3$\pm$0.5\,mas, and an angular size
for the largest component of 0.44$\pm$0.05\,mas (17$\pm$4\,R$_\odot$ at
a distance estimated to be 350$\pm$50\,pc). 

The orbital parameters are shown in Table \ref{tableParameters}. They
are extracted from the spectroscopic and spectrophotometric
observations in \citet{1997A&A...328..219S},
\citet{1999A&A...345..163D}.The inclination derived by
\citet{2000A&A...358..187D} of 63$\pm$3 degree based on the refinement
of their model (O and WR V magnitude) is not taken into account in
this study because the error bar is probably underestimated and relies
on the Hipparcos distance that may not be
correct. \citet{1997A&A...328..219S} performed a new fit of the
polarisation data of \citet{1987ApJ...322..870S} but the uncertainty
for their $\Omega$ parameter is not given in the paper. We adopt the
standard deviation ($\pm$11\degr) of the position angle of the linear
polarization vector provided in \citet{1987ApJ...322..870S}. 
Note that this large error may be partially explained by intrinsic
variations of the polarization due to the wind-wind collision
\citep[see for instance ][]{2005ApJ...623.1092V}. We also stress
that the angle $\Omega$ in the polarimetric model of
\citet{1982MNRAS.198..787}, used in \citet{1987ApJ...322..870S} and
\citet{1997A&A...328..219S} denotes the angle between the North and
the projection of the rotation axis (orbit normal) on the plane of the
sky. This definition do not coincide with the usual definition of the
$\Omega$ parameter used to denote the line of node of binary orbits,
but is flipped by 90\degr.


The combination of the projected orbital radius $a\sin i$ with the
inclination and the distance yields the angular semi-major axis of
the relative orbit $a=4.8\pm$0.7\,$mas$.

We used the \citet{1997A&A...328..219S} ephemeris for
$\gamma^2$~Velorum to calculate the orbital phase at which the
observations were performed (see table \ref{tableParameters}). From
the time of periastron passage T$_0$=2,450,119.1 (HJD), and orbital
period P=78.53 days, the periastron occurs at zero phase, the O-type
component is in front shortly afterwards at phase $\Phi = 0.03$ and
the WR is in front at phase $\Phi = 0.61$. For the date of the VLTI
observation (T=2,453,365.16 (HJD)), using this ephemeris and the
adopted orbital elements, we find an orbital phase $\Phi =
0.32\pm0.03$, i.e., close to quadrature, and we determine the relative
position of the components on the sky. The angular separation of the
stars should be 5.1$\pm$0.9\,mas with a position angle of
66$\pm$15\degr.

This separation is close to the fringe spacing provided by the
baselines. Hence we do not expect to see a fast modulation of the
visibility through the wavelength range considered. Nevertheless, the
visibility signal changes rapidly between the three different projected
baselines of the triplet and as the projected baselines move with the
earth rotation.

\begin{table}[htbp]
  \caption{
    \footnotesize{
      Parameters of the system from the studies of
      \citet{1997A&A...328..219S}.}
    \label{tableParameters}
  }

  \begin{tabular}{lcc}
    \hline
    Parameter & Value & Error\\

    \hline \\

    Distance & 258 pc & +41/-31\\
    Period & 78.53 day & 0.01\\
    Periastron & 2450120.5 day & 2\\
    Eccentricity & 0.326 & 0.01\\
    Periastron longitude $\omega_{WR}$ & 68\degr & 4\\
    a1 sini & 39.$10^6$ km & 2.$10^6$\\
    a2 sini & 125.$10^6$ km & 2.$10^6$\\

    \hline \\

    inclination i & 65\degr & 8\\
    PA of node $\Omega$ & 232\degr & 11\\
    R$_{\rm O}$ & 12.4 R$_\odot$ & 1.7\\
    R$_{\rm c}$ of WR star & 3.0 R$_\odot$ & 0.5\\

    \hline \\

    $\theta_{\rm O}$ & 0.48 mas & 0.09\\
    $\theta_{\rm c}$ of WR star & 0.11 mas & 0.06\\
    $\theta_{(\tau_K = 1)}$ of WR star & 0.28 mas & 0.1\\
    $\pi_{\rm (a1+a2)}$ & 4.8 mas & 0.7\\

    \hline
  \end{tabular}
\end{table}

\subsection{A model for the individual spectra}
\label{subsection:modelSpectra}

Short of performing the full radiation-hydrodynamics problem for the
$\gamma^2$~Velorum system, including the radiation field and force
stemming from each stellar component, the optical-depth effects caused
by their winds, and the emission from the hot collision zone
separating them, we limit ourselves, in this section, to the detailed
modeling of the WR and O star fluxes, with the aim of simulating the
interferometric signals from these two sources alone.

The model atmosphere computations are carried out with the code CMFGEN
\citep{1998ApJ...496..407H}, originally designed to model the
expanding outflows of WR stars. CMFGEN solves the radiative transfer
equation in the comoving frame, under the constraint of radiative and
statistical equilibrium, assuming spherical symmetry and a steady
state, and is capable of handling line and continuum formation, both
in regions of small and high velocities (compared to the thermal
velocity of ions and electrons). Hence, it can solve the radiative
transfer problem both for O stars, in which the formation regions for
lines and continuum extend from the hydrostatic layers out to the
supersonic regions of the wind, and for WR stars where line and
continuum both emerge from regions of the wind that may have reached
half its asymptotic velocity.

The $\gamma^2$~Velorum system has been studied in detail by
\citet{1999A&A...345..163D} and \citet{2000A&A...358..187D}. For the
WR component, we start from a model for WR\,135
\citep{2000MNRAS.315..407D} and adjust the parameters to those of de
\citet{2000A&A...358..187D}. Our WR model parameters are: $L_{\ast} =
10^5 L_{\odot}$, \mdot = 10$^{-5}$\msunyr, a volume filling factor of
10\% that introduces a clumping of the wind at velocities above
$\sim$100\,\kms, C/He = 0.15, and O/He = 0.03 (abundances are given by
number). The velocity law adopted allows a two-stage acceleration,
first a fast acceleration up to a velocity $v_{\rm ext}=1100$\,\kms
(characterized by a velocity exponent $\beta_1=1$) and a more extended
slow acceleration at larger radii (velocities) up to the asymptotic
velocity of $\vinfty=1550$\,\kms ($\beta_2=20$; see
\citet{1998ApJ...496..407H} for details and their eq. (8) for the
velocity law). We associate the stellar surface with the layer where
the Rosseland optical depth is $\sim$20. While
\citet{2000A&A...358..187D} obtained $T_{\ast}=57$\,kK (and
$R_{\ast}=3.2$\,\rsun), we find that the near-IR range can be better
fitted by adopting a slightly hotter stellar temperature, i.e.,
$T_{\ast}=70$\,kK (and $R_{\ast}=2.2$\,\rsun). The higher-temperature
model leads to a better match of the near-IR C{\sc iv}/C{\sc iii}
features, while leaving the optical range still well fitted - only the
He{\sc ii}\,4686\AA, the C{\sc iii}\,5696\AA, and the C{\sc
  iv}\,5808\AA\ are noticeably affected but still satisfactorily
fitted. The general appearance of C{\sc iv} and C{\sc iii} in the
AMBER spectra is somewhat smoother than in the model, but the
absorption at 2.05\,$\mu$m is perfectly fitted. We note that a line is
observed at 2.138\,$\mu$m not taken into account in our WR model. This
line is not an artifact since it is also detected in
the visibilities and phases. We employ both models for the
interferometric study described below.

For the O-star model (computed with CMFGEN, see
\citet{2005A&A...436.1049M}), we select an O8.5\,{\sc iii} spectral
type \citep{1999A&A...345..163D} and adopt the spectral distribution
from \citet{2005A&A...436.1049M}\footnote{A database of O-star spectra
  is available at \texttt{http://www.mpe.mpg.de/$\sim$martins/SED.html}}. The
corresponding O-star parameters are $L_{\ast} = 1.8 \times 10^5
L_{\odot}$, \mdot = 4 $\times 10^{-7}$\msunyr, $\vinfty=2240$\,\kms,
$\log g=3.6$, and $T_{\ast}=32.5$\,kK. Other models of O stars were
also tested providing some input on the sensitivity of the AMBER data
to the O star spectral type.

In our analysis below, we scale both spectra using the
$\gamma^2$~Velorum Hipparcos distance of 258pc, and convolve them with
the AMBER instrumental function to provide a spectral resolution of
$\sim$1500. Moreover, the very low reddening to the $\gamma^2$~Velorum
system \citep{1996A&A...315L.193V} leads to no noticeable extinction
in the near-IR and is thus neglected. Following
\citet{2000A&A...358..187D}, we expect a flux ratio between the WR and
the O star of 0.8--1 in the near-IR spectral region covered by
AMBER. This "free" parameter can also be inferred from our AMBER
observations.


\begin{table}[htbp]
  \centering
  \caption{
    \footnotesize{
      Line and continuum formation regions corresponding to the WR\,11
      model, limited to the near-IR range. For each line, we give the
      radius of the peak emission (in R$_{\ast}=3\,R_{\odot})$) and
      that of the maximum flux in the line, normalized to the
      continuum (in brackets). The apparent diameters are scaled to a
      distance of 258\,pc.}
    \label{tableRadii}
  }
  \begin{tabular}{lcc}
    \hline

    Parameter & Radius in $R_{\ast}$ &
    UD (mas)\\

    \hline\\

    $R_{\tau(2.0\mu m)} $ & 2.7 & 0.30\\
    $R_{\tau(2.5\mu m)} $ & 3.3 & 0.37\\
    C{\sc iv} (2.07\,$\mu$m)  & 3-7 (2.5) & 0.46\\
    C{\sc iii}/He{\sc i} (2.11\,$\mu$m)  & 5-10 (1.4) & 0.57\\
    He{\sc ii}/He{\sc i} (2.17\,$\mu$m)  & 4 (1.2) & 0.34\\
    He{\sc ii} (2.19\,$\mu$m)  & 4 (1.3) &
    0.35\\

    \hline

  \end{tabular}
\end{table}

WR outflows are optically thick up to a few stellar radii above the
hydrostatic surface. The denser the wind, the larger the radius of the
effective photosphere where photons escape, and the more so at longer
wavelengths due to the increase in free-free opacity. Table~\ref{tableRadii} 
lists the radius where the inward integrated continuum optical depth
reaches unity for a range of near-IR wavelength (comparable for both
WR models): for a core radius of $\sim$3\,\rsun, this extends from
1.8 to 3.3\,$R_{\ast}$, from 1 to 2.5\,$\mu$m, equivalent to
$\sim$0.27\,mas. We adopted the Hipparcos distance bearing in mind
that the uncertainty on the star radii can be important as a
consequence of the distance one (see discussion).

\begin{figure}[htbp]
  \centering
  \includegraphics[height=0.33\textwidth, angle=0]{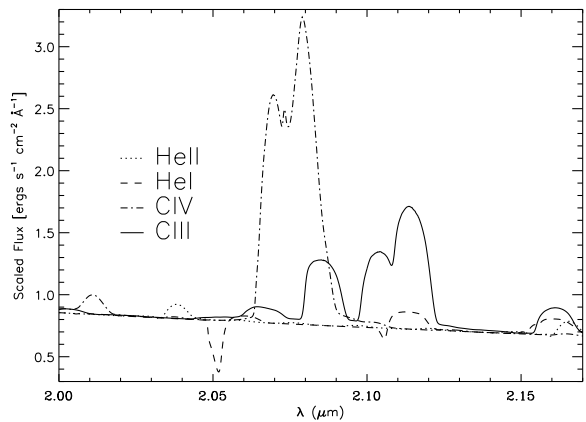}\\
  \caption{
    \footnotesize{
      Near-IR synthetic spectra computed by accounting for
      bound-bound transition of selected ions (He{\sc i}: dashed
      line; He{\sc ii}: dotted line; C{\sc iii}: solid lines; C{\sc
        iv}: dash-dotted line), illustrating the different line
      contributions and overlap in the AMBER spectral windows.
    }
  }
  \label{figure:LFR}
\end{figure}

Photons falling in spectral regions where they experience line as
well as continuum opacity will escape at still larger radii than
photons experiencing exclusively continuum opacity.




More generally, in the 2-2.2\,$\mu$m region, the C\,{\sc
  iii}/C\,{\sc iv}/He\,{\sc ii} lines form over a region exterior to
the (local) continuum photosphere that extends out only to about a
factor of two in radius, corresponding to an angular size of
$\la$0.7\,mas (see table~\ref{tableRadii}). In the $K$ band, stars with
angular diameter below 1\,mas are only marginally resolved with a
baseline of 100\,m. Thus, the variations in WR diameter quoted here
are a second-order signal ($\Delta V \leq 2$\%) difficult to extract
with the current performances of AMBER instrument. Observations with
longer baselines are therefore needed to investigate this particular
point
\footnote{Note that the model also account for the large angular
  diameter found by Hanbury Brown et al. (1970) in the line C{\sc
    iii}\,0.46\,$\mu$m. Its formation region is one of the most
  extended of all optical and near-IR lines.}.

Assuming a flux ratio of about unity in the continuum and that the two
components are essentially unresolved by AMBER, we expect a contrast of
unity for the binary modulation (we also assume in this case the
absence of other contributions in $K$ band from dust and/or the
WWCZ). Thus, given the slowly changing {\it continuum} flux ratio between
the WR- and O-star components, the presence of lines is expected to lead
to a sudden change of the AMBER interferometric signal.


\section{Analysis of the interferometric data}
\label{section:analysis}

\subsection{Analytical fit using visibilities and closure phase}
\label{subsection:analytical_fit}

\begin{table*}[htbp]
  \centering
  \caption{
    \footnotesize{
      Selected continuum zones as in Fig.~\ref{figure:Spectrum} and the
      adjustment of the binary star model. It shows globally a
      constant separation and position angle of 3.65 mas and
      72.7\degr. We have redundant measures separated by 15mn in time
      at 2.025\,$\mu$m and 2.098\,$\mu$m since the spectral windows
      overlapped at this continuum zone. The RMS column corresponds to
      the standard deviation between all the measurements whereas the
      $\Delta$ one represents the average error on the parameters from
      the fitting process.
    }
  }
  \label{table_Fit}
  \begin{tabular}{lccccccccccc}
    \hline
    Wavelength ($\mu$m) & 1.985 & 2.025$_1$ & 2.025$_2$ &
    2.045 & 2.098$_1$ & 2.098$_2$ & 2.1275 &
    2.150 & Avg. & RMS & $\Delta$\\

    \hline\\

    \multicolumn{12}{c}{Observation}\\

    \hline

    Visibility UT2-UT3 & 0.50 & 0.52 & 0.47 &
    0.54 & 0.54 & 0.58 & 0.59 &
    0.59 & & \\

    Visibility UT3-UT4 & 0.26 & 0.26 & 0.28 &
    0.28 & 0.23 & 0.20 & 0.24 &
    0.26 & & \\

    Visibility UT2-UT4 & 0.44 & 0.43 & 0.42 &
    0.42 & 0.41 & 0.53 & 0.52 &
    0.48 & & \\

    Closure Phase & 1.09 & 1.22 & 1.11 &
    1.37 & 1.49 & 1.61 & 1.78 &
    1.60 & & \\

    \hline\\

    \multicolumn{12}{c}{Fit Binary}\\

    \hline

    Separation (mas) & 3.69 & 3.68 & 3.57 &
    3.53 & 3.60 & 3.72 & 3.69 &
    3.68 & 3.65 & 0.06 & 0.1 \\

    Position Angle (\degr) & 64.0 & 64.8 & 93.4 &
    68.8 & 76.4 & 72.3 & 71.1 &
    70.8 & 72.7 & 8.7 & 10\\

    Flux Ratio 2nd star & 0.57 & 0.57 & 0.57 &
    0.59 & 0.64 & 0.68 & 0.66 &
    0.64 & 0.62 & 0.04  & 0.1\\

    \hline
  \end{tabular}
\end{table*}

The aim of this section is to infer the geometrical parameters of
the binary using absolute interferometric observables
exclusively. Specifically, we seek the separation $\rho$, the
position angle $\theta$ of the system, and the flux ratio $R$
between the two components. We perform the fit in the continuum
regions defined in Fig.~\ref{figure:Spectrum}.

\subsubsection{Method and results}
\label{section:anaFitResult}

The geometrical model used to fit the data is a standard binary
model characterized by the astrometric parameters (position angle,
separation, used in the vector $\overrightarrow{\rho}$) and the flux
ratio between the two stars, i.e., $R(\lambda)$, at a given
spatial frequency $\overrightarrow{u_{jk}}$ (see
Eq.~\ref{eqn_Star}),

\begin{equation}
  C_{jk}(\lambda) = \frac{1 + R(\lambda) e^{-2i \pi
      \overrightarrow{u_{jk}} \cdot \overrightarrow{\rho} } }{1 +
    R(\lambda) }\,.
  \label{eqn_Star}
\end{equation}

We use this complex visibility to compute the differential
visibility and phase. Note that at this stage, the algorithm does
not allow to disentangle the irrespective O- and WR-star fluxes. The
absolute visibility is $V_{jk}(\lambda) = |C_{jk}(\lambda)|$, and
the closure phase is $\psi_{123}(\lambda) = \arctan \left[
  C_{12}(\lambda) C_{23}(\lambda) C_{13}^*(\lambda)\right]$.

We used the following set of observables to perform the fit:

\begin{itemize}
\item The absolute visibility for each selected wavelength.
\item The closure phase for each selected wavelength.
\end{itemize}

The binary parameters obtained with this method are shown in the
table.~\ref{table_Fit}. The resulting separation is
3.65$\pm$0.12\,mas. The error bar is computed as follows: we have a
standard deviation between all the measurements of 0.06\,mas, and we
have an average estimate by the $\chi^2$ of the fit to
0.1\,mas. Hence the error is on the order of $\pm$0.12mas. The
position angle is 73$\pm$13\degr, and the flux ratio between the
two stars is 0.62$\pm$0.11.

\subsubsection{Interpretation}
\label{section:anaFitInterpret}

The main point is that the obtained separation is not in a
good agreement with the expected one 5.1$\pm$0.9\,mas. On the
  other hand, the expected position angle of 66$\pm$15\degr is
  compatible within the error bars with the measured one. This result
  is tested in the following sections and discussed in
  Sect.~\ref{section:discussion}. 

We can notice a correlation between the flux ratio and the
wavelength. Such a correlation could be explained if the O
star is the primary and the WR star is the secondary. However, the
estimated flux ratio variation with wavelength is too strong to be
explained by this way. Considering the error bars we have, we can
only say that this trend is fortuitous and that the flux ratio may
be constant over the spectral bandwidth.

We point out to the reader the fact that the flux ratio given by
this method is {\it within} the WR star continuum zones shown in
Fig.~\ref{figure:Spectrum}, which means that the average flux ratio on
all the bandwidth is slightly different. It corresponds in fact to
0.79$\pm$0.12 in average on the 1.95-2.17\,$\mu$m range.


The quality of the fit is not good, the fitted visibilities and
closure phase are on average at 2$\sigma$ over the observed ones. This
overestimate of visibilities means that the ``real'' observed object
is more resolved than a binary star alone.

The easiest way to improve the fit is to consider a third component
for the flux that would be fully spatially resolved and would dilute
the correlated flux observed by AMBER. According to the visibilities,
this contribution could contribute up to 20\% of the overall flux
of the system. However, this significant flux would have been detected
by other techniques and is not reported in the literature. This may
also mean that the observed absolute visibilities on both the
UT2-UT3 and UT2-UT4 bases are significantly biased. This may be
related to the observation problems we noticed on UT2.

\begin{figure*}[htbp]
  \centering
  \begin{tabular}{cc}
    \includegraphics[width=0.48\textwidth, angle=0]{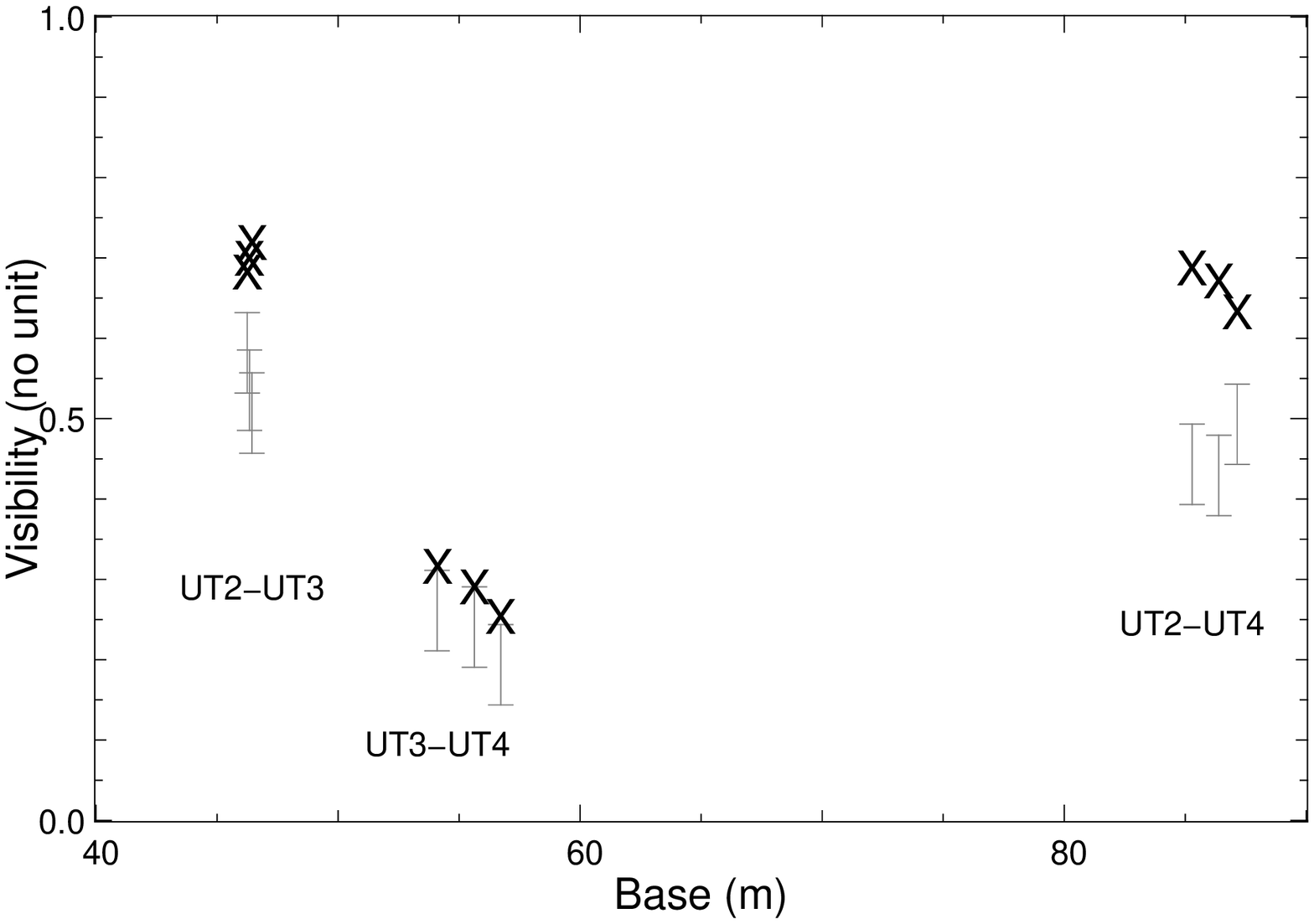}&
    \includegraphics[width=0.48\textwidth,
    angle=0]{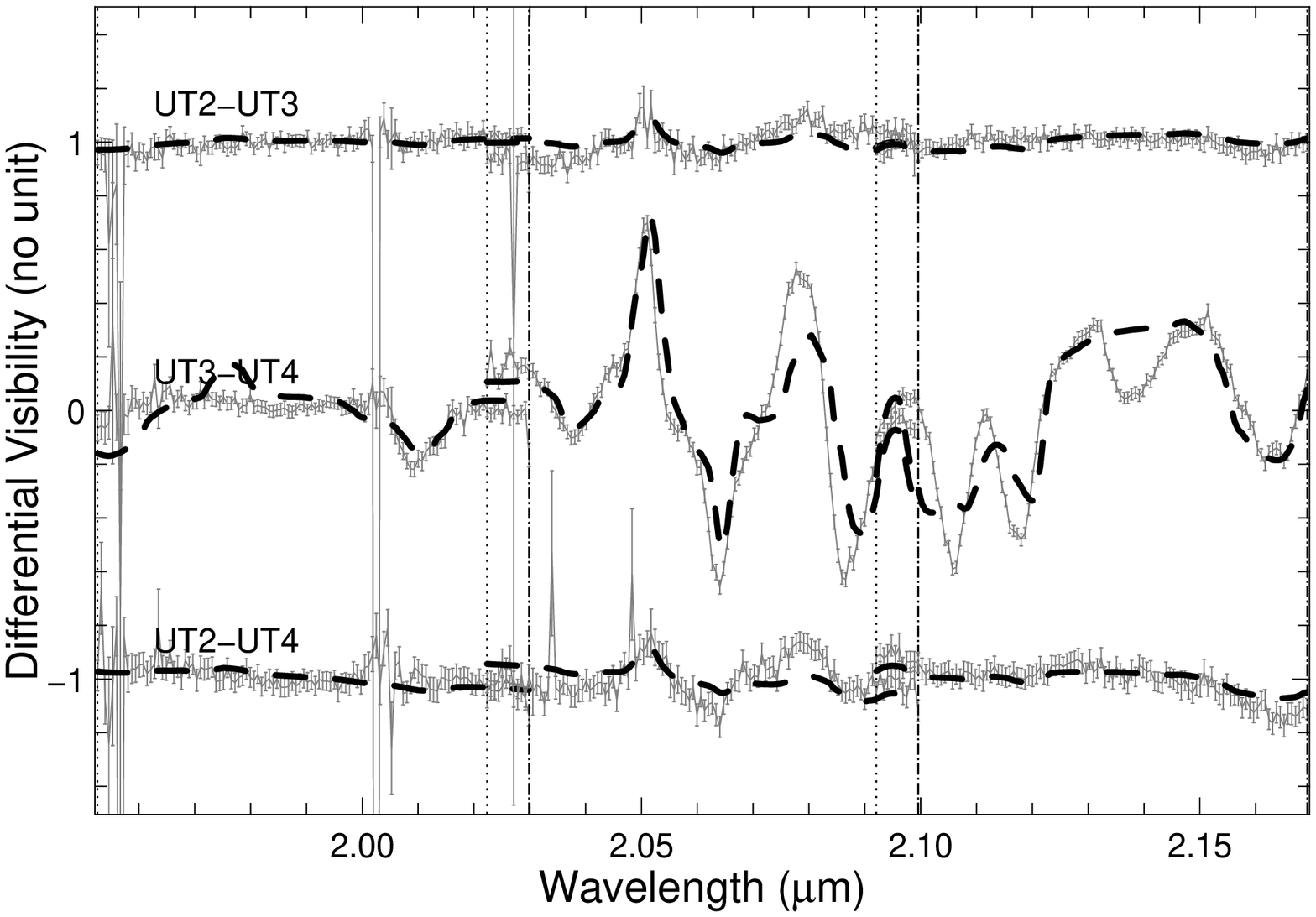}\\
    \includegraphics[width=0.48\textwidth,
    angle=0]{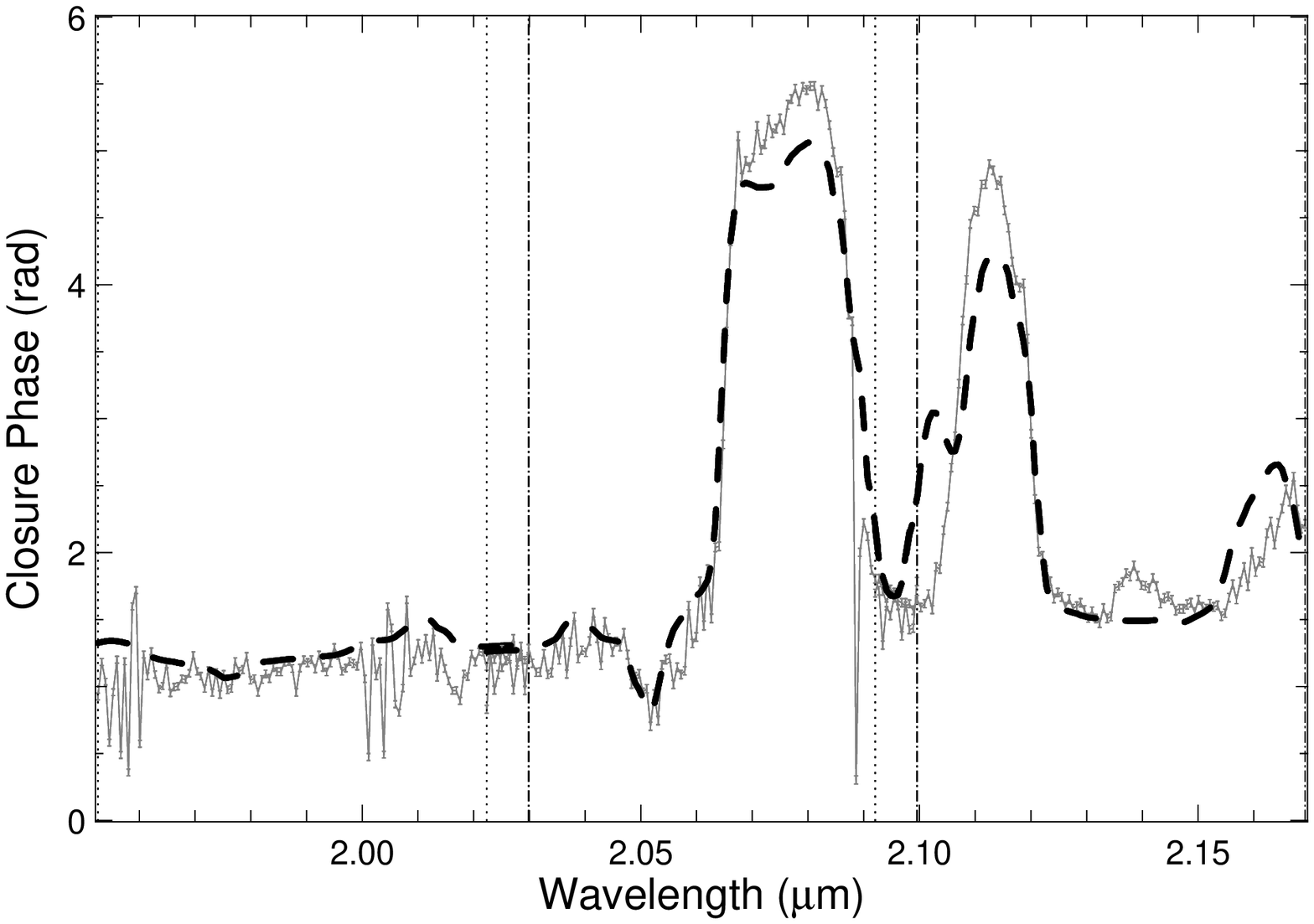}&
    \includegraphics[width=0.48\textwidth,
    angle=0]{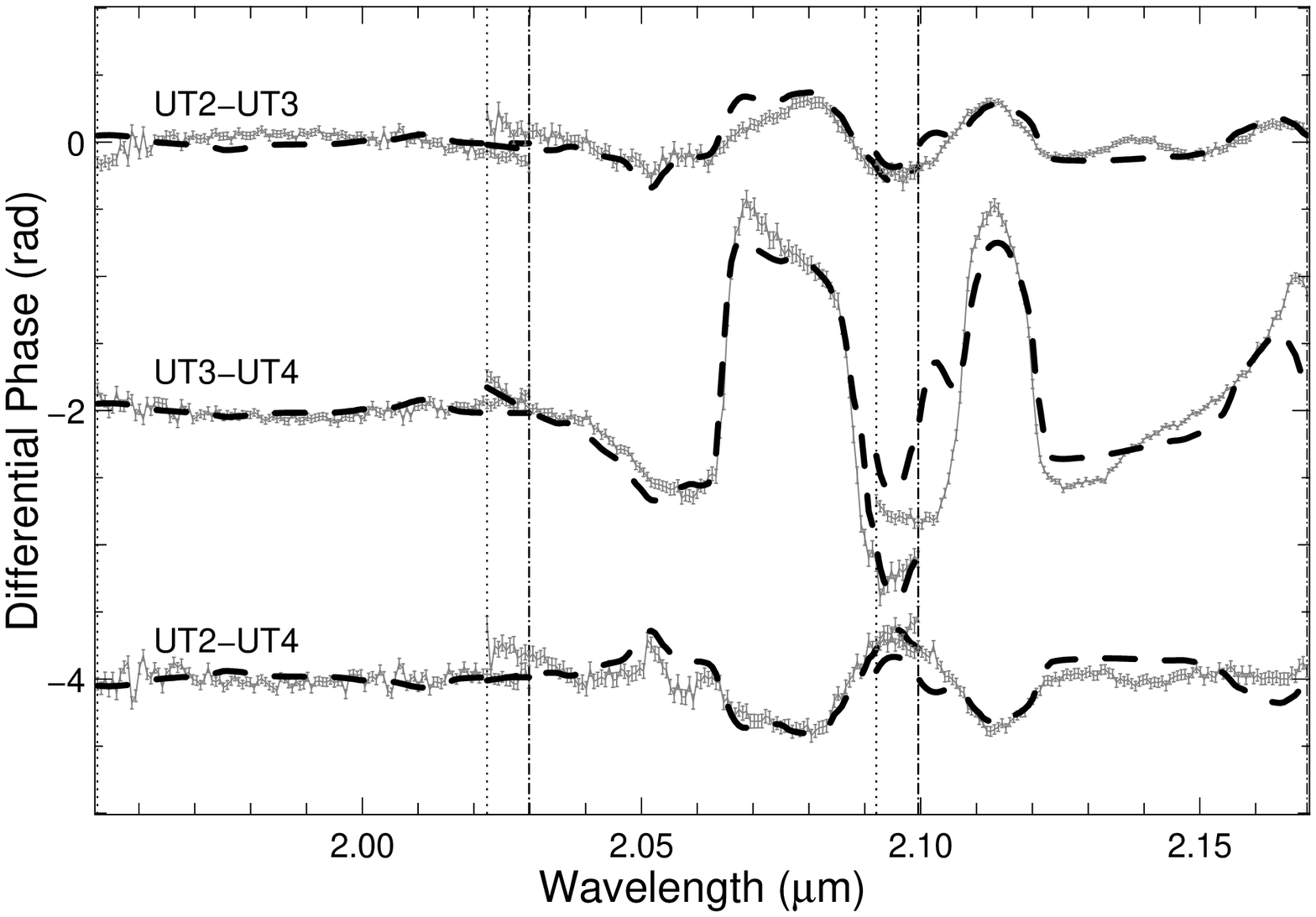}\\
  \end{tabular}
  \caption{
    \footnotesize{
      Observed data with AMBER and the best fit, using a geometrical
      model of a double star and a O- and WR-star synthetic spectra
      of Sect.~\ref{subsubsection:MODEL}.
      {\it top-left}:
      Points with error bars: observed absolute visibilities versus
      base length. Crosses: our model. See text for comments.
      {\it top-right}:
      Gray line with error bars: observed differential visibilities
      versus wavelength. Dashed line: our model. The different
      baselines are offseted for clarity.
      {\it bottom-left}:
      Gray line with error bars: observed closure phase versus
      wavelength. Dashed line: the model. See text for comments.
      {\it bottom-right}:
      Gray line with error bars: observed differential phases versus
      wavelength. Dashed line: our model. The different baselines
      are offseted for clarity.
    }
  }
  \label{figure:MODEL}
\end{figure*}

\subsection{Fit using WR and O stars modeled spectra}
\label{subsubsection:MODEL}


In this section we use the synthetic spectra presented in 
Sect.~\ref{subsection:modelSpectra} in order to find the  
parameters that best match our data. We still assume the
individual components of the binary system are unresolved. 

\subsubsection{Method and results}
\label{section:MODELResult}


In Sect.~\ref{subsection:analytical_fit}, we restricted the fitting
process to a selection of a few narrow continuum windows. We now
wish to perform a fit using the information from the full spectral
window (about half the $K$ band). In order to perform such a fit, we
use the geometrical model described in Eq.~\ref{eqn_Star}, together
with the synthetic spectra described in
Sect.~\ref{subsection:modelSpectra} \citep{2005A&A...436.1049M}. We
are particularly interested in seeing a change in the
interferometric signal associated with the predicted change of the O
to WR flux ratio as we progress from continuum to line regions.

The set of observables considered to perform the fit has been
extended to the full dataset, namely the spectrum, the averaged
absolute visibilities and closure phase per spectral window, the
differential visibilities and the differential phases.

We use non-linear fit methods in order to minimise the $\chi^2$
between the observables and the model. Then we compute the best fit of
about a thousand randomly chosen initial parameters in order to get
the best minimum of $\chi^2$. The final model has then been compared
to the observed star spectrum and interferometric observables as
shown in the Fig.~\ref{figure:MODEL}.

The best fit yields a binary star separation of
3.64$^{+0.09}_{-0.40}$\,mas, a position angle of
72$^{+17}_{-14}$\degr, a flux ratio of 0.75$^{+0.10}_{-0.08}$ (in
the whole 1.95-2.17\,$\mu$m range), attributed here to the WR to O
flux ratio. This detection is made possible because of the non-zero
closure phase signal and is clearly made because of the presence of
different lines in the WR- and O-star spectra.


%

The AMBER instrument would normally allow to disentangle definetly
which component is the North-East and which is the
South-West. However, the calibration data obtained for this has been
taken and is still in the process of being interpreted. This is why we
guess, according to a preliminary study of this calibration data, that
the North-East component is the WR star and the South-West is the O
star, but we are quite prudent about this particular point.


\subsubsection{Interpretation}
\label{section:MODELInterpretation}

We tested the fits with the full library of spectrum provided by
Martins et al. (2005). The quality of the fits is only slightly
affected by the choice of the O star spectrum. In the near-IR, the
spectrum is weakly sensitive to the star temperature and equally
good fits can be obtained with models with T$_{\rm eff}$ between
27kK and 35kK (or yet higher). Yet, the four spectra providing the
best fits are those with log$g$ between 3.2 and 3.35, considering
the O star as a supergiant. This information has to be taken with
caution since the residuals depend critically on the choice of the
WR star spectrum, but this result still holds when we consider our
different WR models.

The spectrum, the differential visibilities, differential phases and
closure phases are reasonably well fitted. The absolute visibilities
are overestimated in our model compared to that of AMBER. Again, these
discrepancies may be due to biases in the absolute visibilities of
AMBER, but we note also significant departures in
the differential visibilities, differential phases and closure phase
at 2.08\,$\mu$m (C{\sc iv} line), 2.115\,$\mu$m (C{\sc iii} line)
and 2.14\,$\mu$m (see Fig.~\ref{figure:MODEL}).

As mentioned in the previous section, the simplest way to solve the
absolute visibilities discrepancies is to add a fully resolved
``continuum'' contribution. However, the constraints provided by the
differential observable and the closure phase are also tight, due to
the large flux ratio variations in the WR lines. This leave only small
room to even a small diluting factor. We tried to inject a fully
resolved component with varying flux contribution and the maximum
possible continuum contribution has been estimated to 5\% of the
overall flux. Within this range, the fits of the differential
observables and the closure phase are slightly improved, but the
discrepancies of the absolute visibilities remain.

The most convincing signature of the WWCZ may be found in lines, but
in the stage of development of the WR spectrum model, it is not
absolutely sure that the residuals of the fits in the lines come from
an inadequacy of the models or an intrinsic signal from an additional
component.

\subsection{WR spectrum reconstruction}
\label{subsubsection:CONTRAINT}


\begin{figure*}[htbp]
  \centering
  \begin{tabular}{cc}
    \includegraphics[width=0.48\textwidth, angle=0]{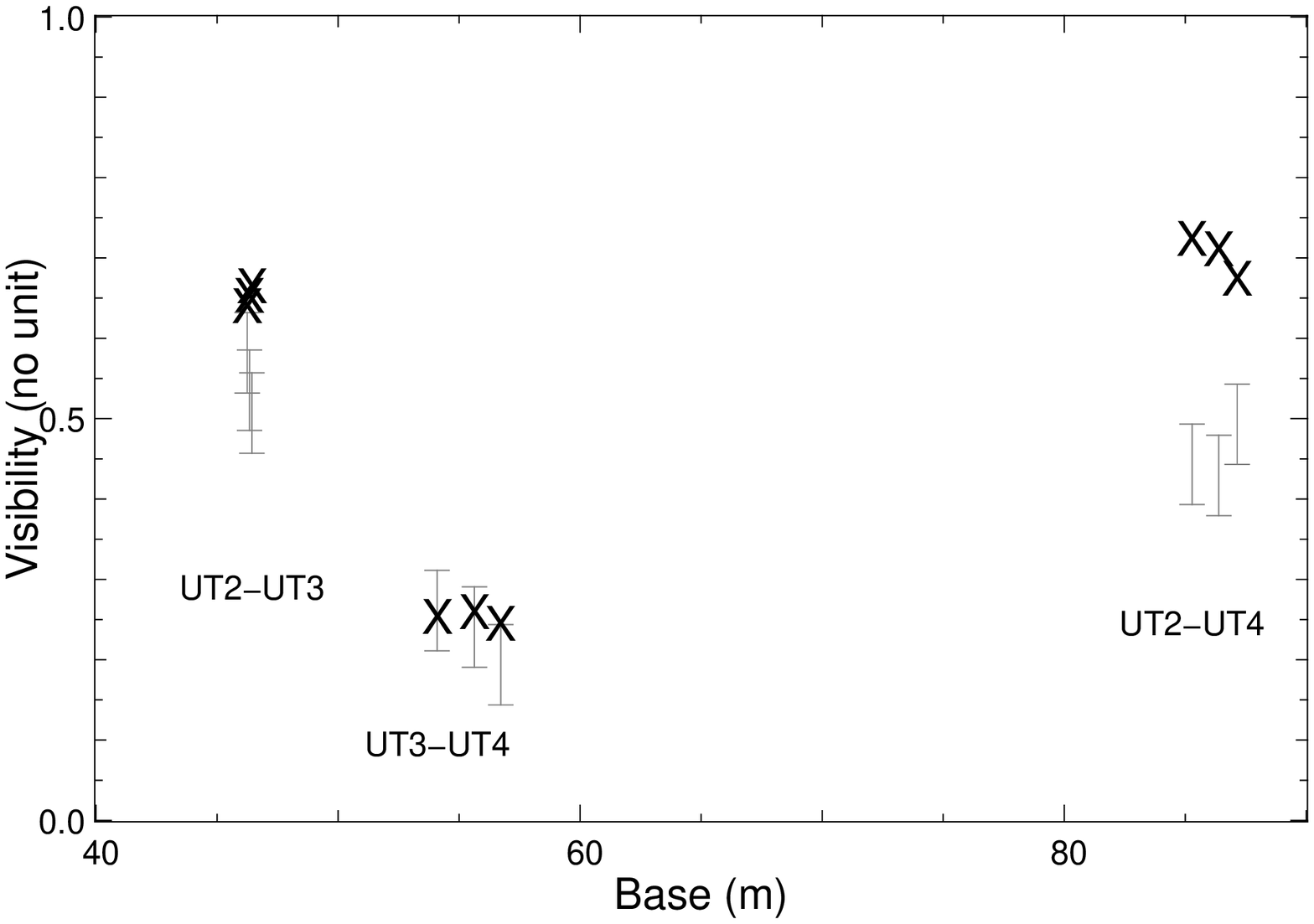}&
    \includegraphics[width=0.48\textwidth,
    angle=0]{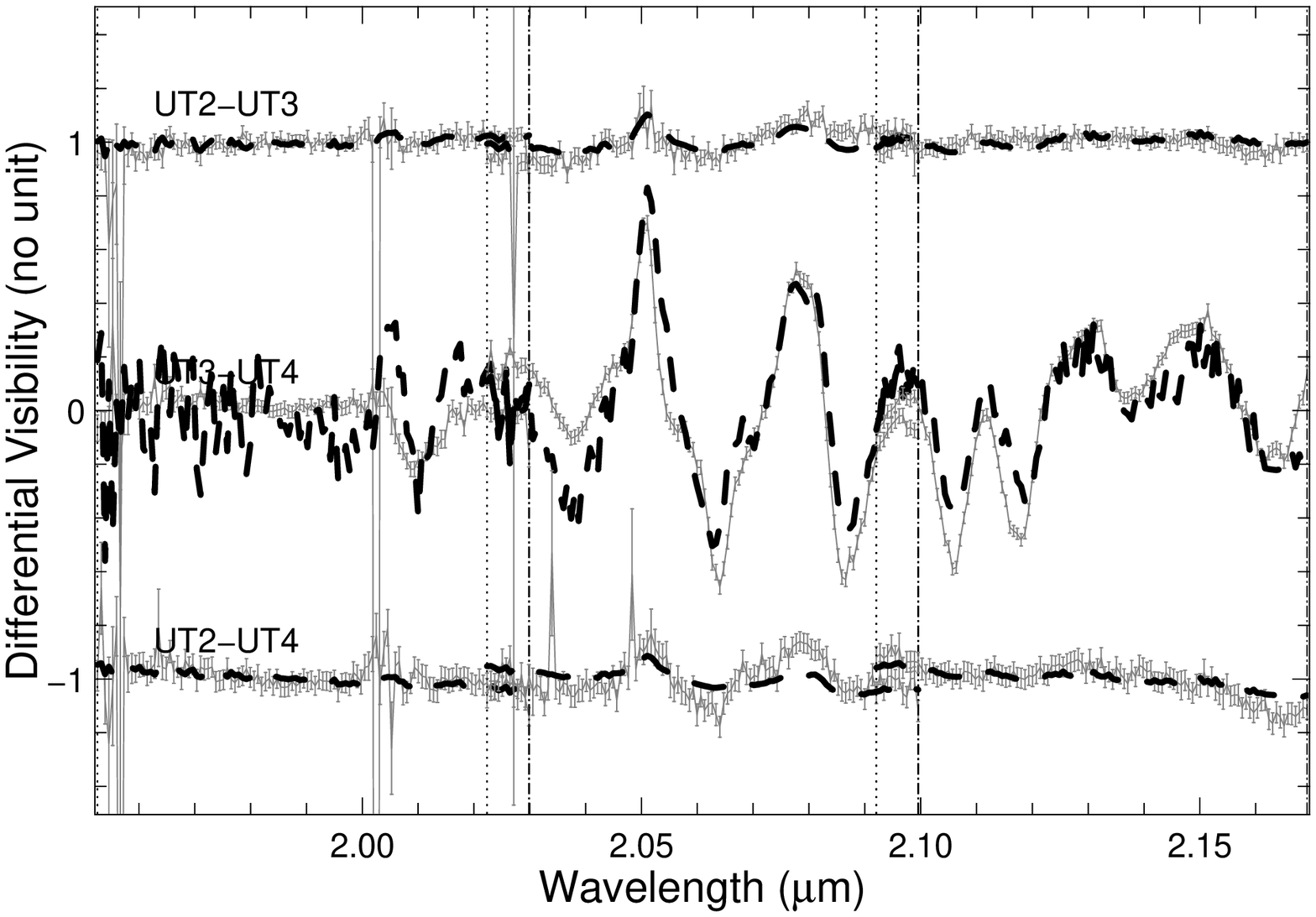}\\
    \includegraphics[width=0.48\textwidth,
    angle=0]{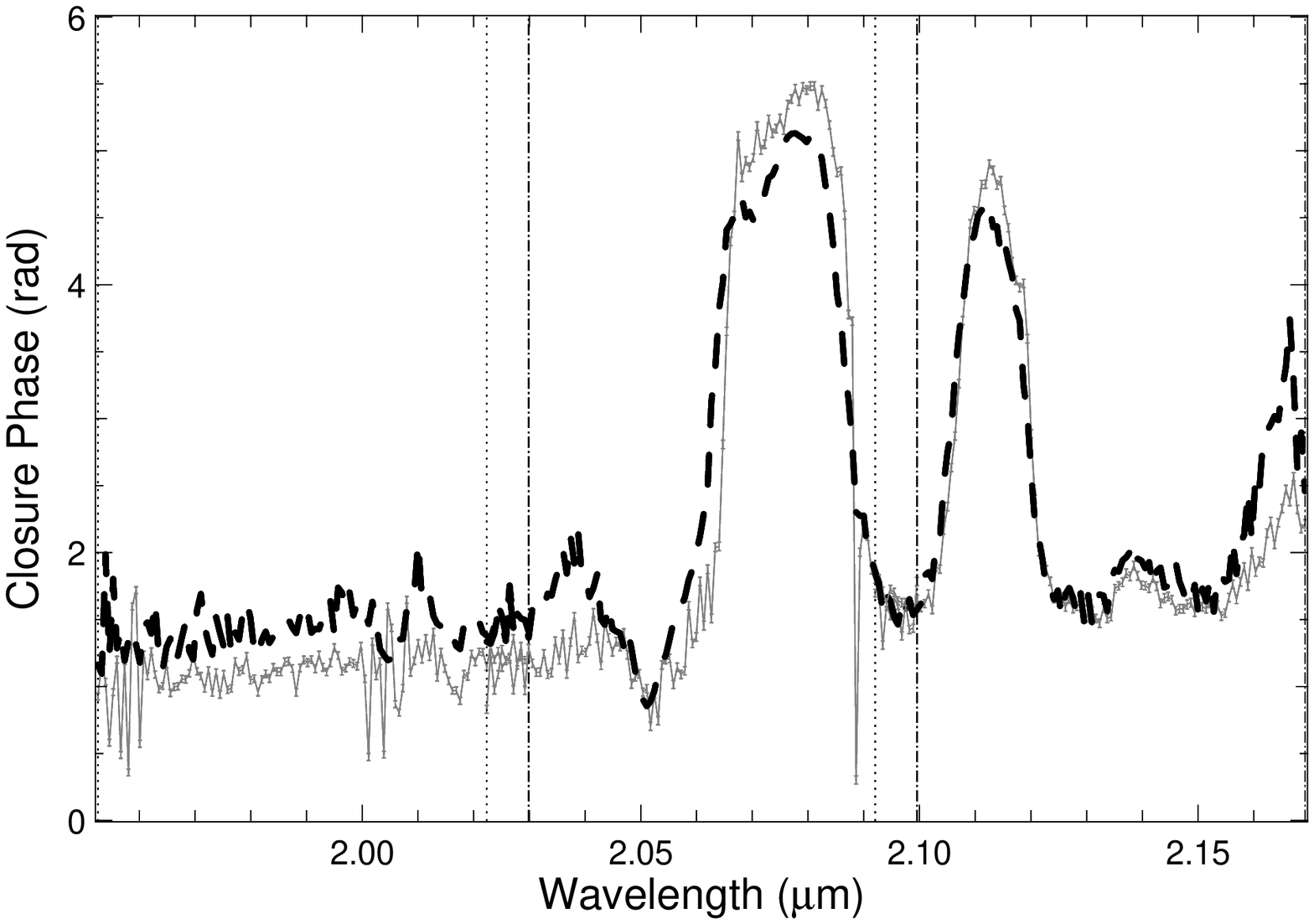}&
    \includegraphics[width=0.48\textwidth,
    angle=0]{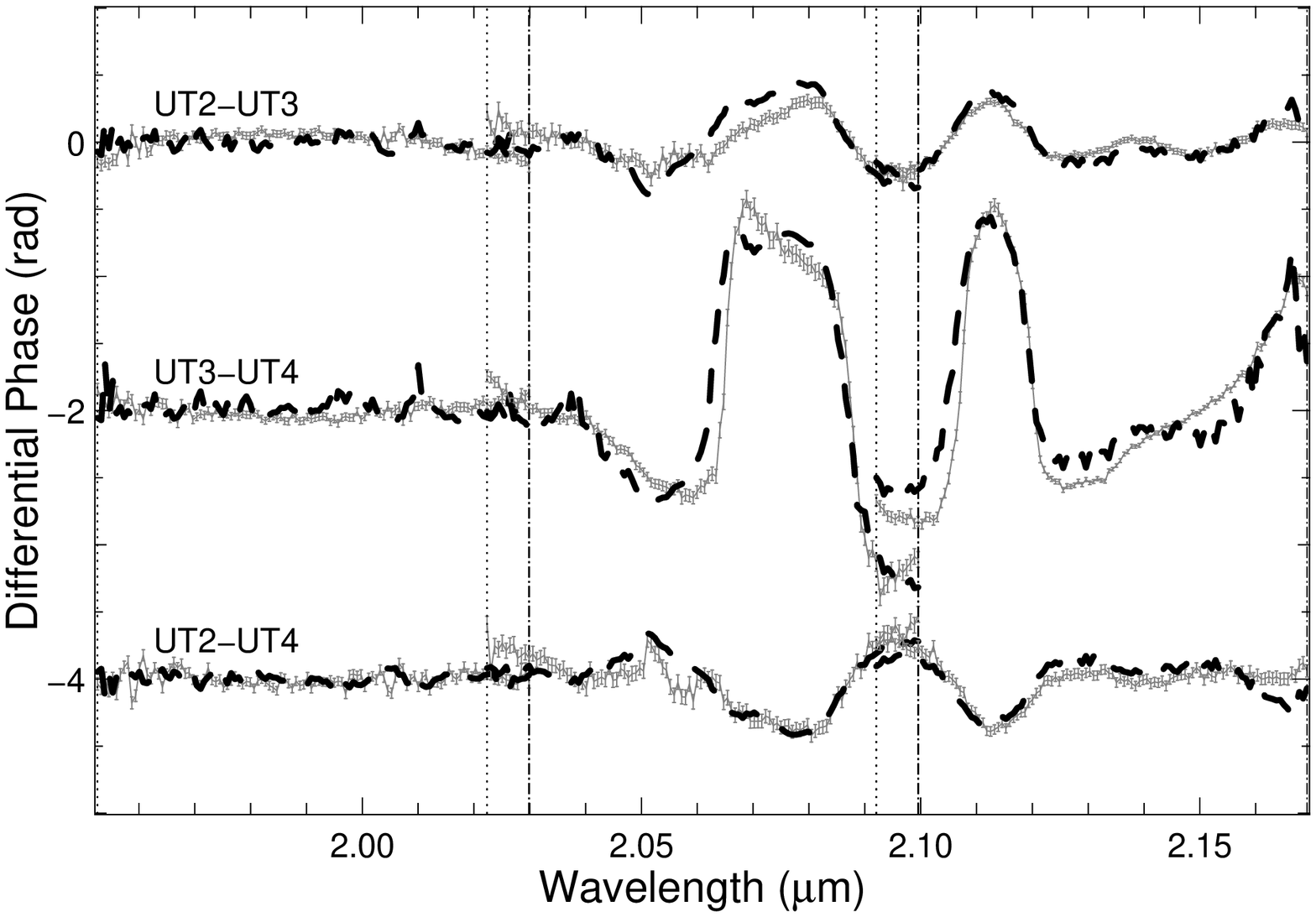}\\
  \end{tabular}
  \caption{
    \footnotesize{
      Observed data with AMBER and the best fit, using a geometrical
      model of a double star, a O-star synthetic spectrum and a
      reconstructed WR-star spectrum of
      Sect.~\ref{subsubsection:CONTRAINT}.
      {\it top-left}:
      Points with error bars: observed absolute visibilities versus
      base length. Crosses: our model. See text for comments.
      {\it top-right}:
      Gray line with error bars: observed differential visibilities
      versus wavelength. Dashed line: our model. The different
      baselines are offseted for clarity.
      {\it bottom-left}:
      Gray line with error bars: observed closure phase versus
      wavelength. Dashed line: the model. See text for comments.
      {\it bottom-right}:
      Gray line with error bars: observed differential phases versus
      wavelength. Dashed line: our model. The different baselines
      are offseted for clarity.
    }
  }
  \label{figure:CONTRAINT}
\end{figure*}

\subsubsection{Method and results}
\label{section:contraintResult}

In this section, we consider that the O star spectrum is better
constrained than the WR star spectrum. It is just an almost
featureless continuum with a relatively well-defined slope. Hence, we
try another approach based on our simple geometrical model of a binary
with unresolved components of Eq.~\ref{eqn_Star}. Previously, the
wavelength-dependent flux ratio between the O star and the WR star was
defined by the ratio of the synthetic spectra.

Now, we determine for each spectral channel the WR star flux using
only the observed flux and the O star model. The idea is to use all
the information contained in the data in order to minimize the {\it a
  priori} information used in the model. We point out to the reader that the observed spectrum is normalized as described in 
Appendix~\ref{section:SpecCalib}, the absolute flux information being
lost.

Let $S^{\rm GV}(\lambda)$ be the observed spectrum, $R$ the flux ratio
between the O star and the WR star, and $S^{\rm O}_{\rm N}(\lambda)$ 
the normalized O spectrum. We can define a normalized WR star spectrum by:

\begin{equation}
  S^{\rm WR}_{\rm N}(\lambda) = \frac{ (1+R) * S^{\rm GV}(\lambda) -
    S^{\rm O}_{\rm N}(\lambda)}{R}
  \label{eqn_SWR}
\end{equation}

Since the observed spectrum is normalized as described in
Appendix~\ref{section:SpecCalib}, we need to normalize in the
same way the O star synthetic spectrum before subtracting it from
the observed spectrum in order to get this normalized WR star
spectrum. Then we multiply the resulting spectrum with the slope of
a blackbody at 56000\,K, the expected temperature of the WR star. We
tested several black body temperatures such as 70000\,K for example
and found that it does not change dramatically the slope of the WR
spectrum nor the result of the fit. This suggests that using a Black
Body for this fit is approximate but adequate since the slope of the
energy distribution of a WR star is not too sensitive to the
temperature of the model in the 1.95-2.17\,$\mu$m range.


At this point we get a completely constrained spectrum of the WR star,
only dependent from the O star model and the AMBER spectrum. We inject
then the spectrum of the modeled O star and the constrained WR star
in the interferometric data in order to compute a $\chi^2$ and perform
the fit.

This technique has been used to fit the data shown in
Fig.~\ref{figure:CONTRAINT}. Compared to the previous method
involving the synthetic WR spectrum, the residuals (i.e. the
$\chi^2$) are smaller. For instance, the contribution from the
missing line at 2.138\,$\mu$m, not predicted by our WR model, is well
reproduced and the residuals are small. The poor photometric quality
of the spectral window between 1.95 and 2.2\,$\mu$m (and in
particular the 2.01\,$\mu$m atmospheric feature) is reported in the
differential visibilities (mostly the UT3-UT4 ones) and the closure
phase.

This method provides the best fit to the data of the present paper and
all the following is based on the results of this fit. It yields the
following parameters: a binary separation of 3.62$^{+0.11}_{-0.30}$\,mas,
a position angle of 73$^{+9}_{-11}$\degr, a WR flux contribution in
the 1.95-2.17\,$\mu$m spectral window of 0.79$^{+0.06}_{-0.12}$.


The parameters are basically unchanged compared to those of the
previous section, suggesting that the quality of the WR synthetic
spectrum does not introduce a sizable bias on our determination.



\section{Discussion}
\label{section:discussion}

\subsection{Spectra separation and stars parameters}

\begin{figure}[htbp]
  \centering
    \includegraphics[width=0.48\textwidth, angle=0]{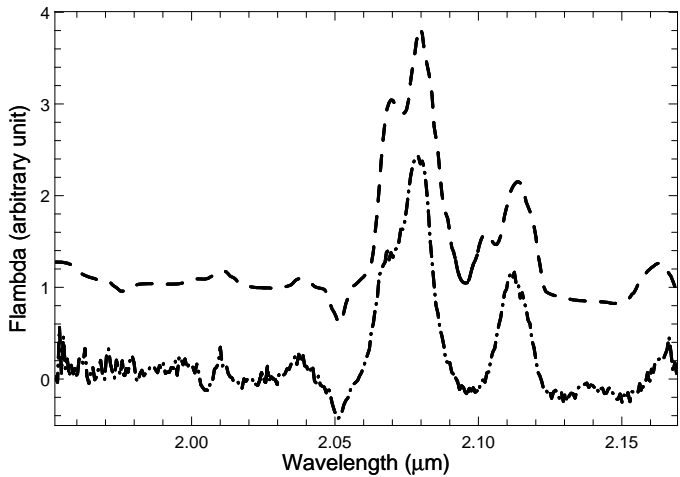}\\
  \caption{
    \footnotesize{
      The two final WR spectra showed on the same plot. Dashed line:
      WR model of Sect.~\ref{subsubsection:MODEL}. Dash-dotted line:
      WR spectrum of Sect.~\ref{subsubsection:CONTRAINT}. They show
      similarities in the lines at 2.059$\mu$m and 2.165$\mu$m but our
      modeled spectrum is less accurate in the carbon lines at
      2.071$\mu$m, 2.079$\mu$m, 2.108$\mu$m and 2.114$\mu$m.
    }
  }
  \label{figure:ComparSpectrums}
\end{figure}

One very interesting point of the methods described above is the fact
that we are able to extract a WR spectrum independent from previous
spectrophotometric measurements. This allows us to compare our best
spectrum model from the fit of the Sect.~\ref{subsubsection:MODEL} and
this independently extracted spectrum. One can find the resulting
  spectra on the Fig.~\ref{figure:ComparSpectrums}.

\subsection{Binary separation and distance}

\begin{figure}[htbp]
  \centering
  \includegraphics[width=0.4\textwidth, angle=0]{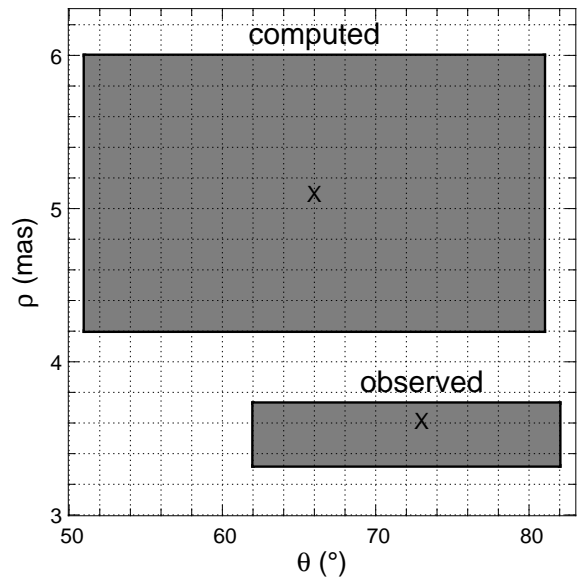}
  \caption{
    \footnotesize{
      Errors derived from the different techniques used to retrieve
      the geometrical parameters of the binary star at the time of the
      AMBER observations. The big
      gray box represents the estimation and error bars from the
      radial velocity method and the small gray box is the resulting
      parameters from our interferometric fit. This figure show that
      the direct measured separation by interferometric means is
      smaller than the expected one, leading to a possible
      reevaluation of the distance of the system.
    }
  }
  \label{figure:Error_boxes}
\end{figure}



The geometrical parameters provided by the different approaches used
to invert the data have been markedly consistent and robust.  The
best fit (i.e. minimum $\chi^2$) and narrower error bars are found
for the method in Sect.~\ref{subsubsection:CONTRAINT} which the WR
spectrum is considered as undefined. This could appear artificial
but the ressemblance of the spectrum and fits estimated with the
ones found using a radiative model lead us to be confident on the
results presented. The parameters are shown in table
\ref{table_results} and the results from
Sect.~\ref{subsubsection:CONTRAINT} are reported in
Fig.~\ref{figure:Error_boxes}. With this direct observation of the
$\gamma^2$~Velorum with the interferometer and our model fitting, we
should be able to disentangle the $\pi$ uncertainty between the
WR and O position, the first being the North-East and the second being
the South-West at the time of the observations (see the
Sect.~\ref{subsubsection:MODEL} for more details). However, this
  preliminary result has to be confirmed with the complete analysis of
  our AMBER calibration data

\begin{table}[htbp]
  \centering
  \caption{
    \footnotesize{
      Summary of all the methods and results used in this paper,
      showing the good agreement we have with several different
      interpretation methods of the interferometric data, but the bad
      agreement on the separation between the interferometric methods
      and the spectrophotometric method.
    }
  }
  \label{table_results}
  \begin{tabular}{lcccccc}
    \hline
    Method & Separation & Pos. angle & Averaged fx. rat.\\
    & (mas)      & (\degr)    & (1.95-2.17\,$\mu$m)\\

    \hline\\

    Sect.~\ref{subsection:geomParamSys} & 5.3$^{+}_{-}$0.9 &
    66$^{+}_{-}$15 & - \\

    \hline\\

    Sect.~\ref{subsection:analytical_fit} & 3.65$^{+}_{-}$0.12 &
    73$^{+}_{-}$13 & 0.79$^{+}_{-}$0.12 \\

    \hline\\

    Sect.~\ref{subsubsection:MODEL} & 3.64$^{+0.09}_{-0.40}$ &
    72$^{+17}_{-14}$ & 0.75$^{+0.10}_{-0.08}$ \\

    \hline\\

    Sect.~\ref{subsubsection:CONTRAINT} & 3.62$^{+0.11}_{-0.30}$ &
    73$^{+9}_{-11}$ & 0.79$^{+0.06}_{-0.12}$ \\

    \hline
  \end{tabular}
\end{table}

The binary parameters remain unchanged within the error bars shown in
the figure and do not change the present reasoning. The estimated
errors on the position angle are relatively large, but comparable to
the uncertainties of the spectroscopic orbit. The values agree well,
which means that the projected position of the two stars is well
predicted if the spectroscopic information is complemented by the
polarization data from \citet{1987ApJ...322..870S}. By contrast, our
estimated projected separation is well-constrained and differs
significantly from the predicted one. The AMBER separation is at a
2$\sigma$ level from the predicted one.



Furthermore, from the AMBER point-of-view, the spectroscopic-based
separation at 1$\sigma$ in Fig.~\ref{figure:Error_boxes} is still
separated to more than 3$\sigma$ of the AMBER error bar. The
spectroscopic-based error bar on the separation $\rho$ is mostly
defined by the Hipparcos uncertainties. A computation of the distance
including the AMBER separation and angle measurements in the frame of
the spectroscopic orbit parameters leads to a distance of
368$^{+38}_{-13}$\,pc.

Before the Hipparcos era, the common estimated distances to
$\gamma^2$~Velorum were typically of $\sim$450~pc
\citep{1988MNRAS.232..821B, 1996MNRAS.283..589S}.  The present
estimate, with a variance at the 2$\sigma$ level from the Hipparcos
measurement, would place $\gamma^2$~Velorum within the Vela OB2
association, affecting all distance dependent parameters such as the
luminosity, radii, and of course spectral types.  As an example, the
Table~\ref{tableParameters} reporting the parameters of
\citet{2000A&A...358..187D} are scaled to the Hipparcos distance. In
pre-Hipparcos area, the spectral type of the O star has been for a
long time O9{\sc i} \citep{1997NewA....2..245V}, which means that the
typical radius is about 20$R_\odot$, rather than 13$R_\odot$.

The reliability of the Hipparcos distance has recently been questioned
by the discovery of an association of low-mass, pre-main sequence
stars in the direction of $\gamma^2$~Velorum which would have affected
the measured parallax, and the distance to $\gamma^2$~Velorum may be
between 360 and 490\,pc \citep{2000MNRAS.313L..23P}. A low-mass
companion 4.8\,$\arcsec$ away has been observed in the X-ray band with
Chandra \citep{2001ApJ...558L.113S}. A similar problem concerning the
star WR\,47 (WN6+O5V) has been reported and extensively studied by
Pietti et al. which leaded to a distance multiplied by 4 compared to
the Hipparcos one (1.10$\pm$0.05 kpc versus 216$^{+166}_{-65}$ pc
\footnote{This distance would have made WR\,47 the closest WR, before
  $\gamma^2$~Velorum.}).

\subsubsection{Residuals of the fits}
\label{section:residuals}

In Sect.~\ref{subsection:modelSpectra}, we have documented the
properties of the $\gamma^2$~Velorum system used to estimate the
observed signal with the AMBER instrument. Our interpretation
suggests a binary system whose separation is resolved by the
interferometer, but not their individual components diameter. Note
that we have neglected the presence of dust or any other  source of
emission.

Different modeling methods were used, with different assumptions on
the two sources, but lead to results in agreement for the binary
separation, the position angle, and flux ratio between the two
stars. However, in both methods, the quality of the fits could be
improved. Errors and biases of the different AMBER observables may
corrupt the fitting process, but the disagreements may also stem from
additional components not yet accounted for.


Let us assess the source of the residuals and try to provide some
information on the way the present model of the $\gamma^2$~Velorum system
could be improved. The residuals of the fits are analyzed per observable:

\begin{itemize}
\item Differential visibilities: The fits based on the extracted WR
  spectrum are of good quality and the residuals can not be
  attributed definitely to an additional astronomical signal. The
  levels of the dips and peaks of the signal are tightly correlated
  to the changing of the primary versus the secondary source of
  flux. A small change of the flux ratio between the two stars, and
  the inclusion of a small source of continuum can affect
  drastically this observable. More complex models, e.g., including
  a continuum contribution of up to 5\% of the total flux, improve
  slightly the match with the observations.
\item Differential phases: the quality of the fits is reasonably
  good considering that this observable is subject to continuum
  fluctuations that can be seen, for instance, in the edge of the
  spectral window at 2.17\,$\mu$m. We note a discrepancy of the
  model and the signal at 2.07\,$\mu$m for the baselines UT2-UT3 and
  UT3-UT4. This discrepancy is only slightly attenuated when using
  the WR synthetic spectrum that shows a stronger contribution at
  this wavelength than observed. This may indicate an additional
  source of emission for this particular line at a position not
  coincident with that of the WR star. The balance of the different
  contributions of these blended lines is temperature dependent and
  the WWCZ can add both a spatial and a spectral signature.
\item Closure phase: there is a noticeable departure of the model in
  the 1.95-2.03\,$\mu$m spectral window. The lines are not fitted
  perfectly but the residuals are at the 10\% level. This may suggest
  a third source in the system, but as before there is an ambiguity in
  the interpretation of this observable and a slight change in the
  shape of the WR lines and/or in the star-flux ratio could also
  change this inferred closure phase.
\item Absolute visibilities: whatever the strategy used to fit the
  AMBER data, the model fails to provide the low level of absolute
  visibilities observed. The obvious solution is the introduction of a
  third source of continuum flux. However the contribution from an
  additional source is strongly limited by the other observables,
  requiring that the modulation of the signal by the WR is sufficient to
  explain the level of spectral variations observed. We tested some
  models with a small amount of (flat) continuum that improved
  slightly the model level but insufficiently. For this observable
  though, we think that some instrumental effect on UT2 biased the
  visibilities.
\end{itemize} 

At this stage of the analysis and given the limited amount of data
available, it is not possible to attribute with confidence the
residuals to an astrophysical origin.  However, we have several
observables that have significant discrepancies in some lines, giving
tenuous clues that something more has to be included in our model to
improve the fits. More data is needed, covering a broad range of
binary phases, in order to get more constraints on a complete
  model.



\section{Conclusion}
\label{section:conclusion}

Using a relatively restrained data set from the AMBER/VLTI instrument,
we have set tight constraints on the geometrical parameters of the
$\gamma^2$~Velorum orbit. This separation leads to a reevaluation of
the distance of the $\gamma^2$~Velorum system that has to be confirmed
and more accurately estimated by a regular monitoring of the system.

Moreover, we were able to perform a spectrum separation between
  the two stars, using known assumptions on the spectral type of the O
  star. This allowed us to compare our modeled WR spectrum to this
  independent one and found that their match is reasonable.


We note however that the observed data set is not fully consistent
with a simple geometrical binary model taking into account refinements
of the modeled spectra for each component. This discrepancy may be
interpreted as not well understood instrumental biases as well as the
detection of a spatially distinct source of continuum, that would
  contribute to up to 5\% of the total flux of the system.


\begin{acknowledgements}
We warmly thank John Davis and Julian North for fruitful discussions on the definition of polarimetric
versus spectroscopic parameters of binary orbits that allowed us to detect an inconsistancy on the direction of the orbit in the paper.

  We thank Fabrice Martins for providing the O star synthetic
  spectrum. This paper makes use of  \emph{Jean-Marie
    Mariotti Center} (JMMC) tools for model fitting and
  data adjustment.

  We would like to thank
  the staff of the European Southern Observatory for
  their help in the design and the commissioning of the AMBER
  instrument.

  This project has benefited from funding from the French Centre
  National de la Recherche Scientifique (CNRS) through the Institut
  National des Sciences de l'Univers (INSU) and its Programmes
  Nationaux (ASHRA, PNPS). The authors from the French laboratories
  would like to thank the successive directors of the INSU/CNRS
  directors.  The authors from the the Arcetri Observatory acknowledge
  partial support from MIUR grants and from INAF grants.
  C. Gil work was supported in part by the Funda\c{c}\~ao para a
  Ci\^encia e a Tecnologia through project POCTI/CTE-AST/55691/2004
  from POCTI, with funds from the European program FEDER.
  
  This research has also made use of the ASPRO observation preparation
  tool from the JMMC in France, the SIMBAD database at CDS, Strasbourg
  (France) and the Smithsonian/NASA Astrophysics Data System
  (ADS). This publication makes use of data products from the Two
  Micron All Sky Survey.

  The AMBER data reduction software \texttt{amdlib} has been linked
  with the open source software
  Yorick\footnote{\texttt{http://yorick.sourceforge.net}} to provide
  the user friendly interface \texttt{ammyorick}. They are freely
  available on the AMBER website
  \texttt{http://amber.obs.ujf-grenoble.fr}.

\end{acknowledgements}

\bibliography{biblio}

\begin{thebibliography}{33}
\expandafter\ifx\csname natexlab\endcsname\relax\def\natexlab#1{#1}\fi

\bibitem[{{Barlow} {et~al.}(1988){Barlow}, {Roche}, \&
  {Aitken}}]{1988MNRAS.232..821B}
{Barlow}, M.~J., {Roche}, P.~F., \& {Aitken}, D.~K. 1988, \mnras, 232, 821

\bibitem[{{Brown} {et~al.}(1982){Brown}, {Aspin}, {Simmons}, \&
  {McLean}}]{1982MNRAS.198..787}
{Brown}, J.~C., {Aspin}, C., {Simmons}, J. F.~L., \& {McLean}, I.~S. 1982,
  \mnras, 198, 787

\bibitem[{{Corcoran} {et~al.}(2003){Corcoran}, {Hamaguchi}, {Henley},
  {Pittard}, {Gull}, {Davidson}, {Swank}, {Petre}, {Ishibashi}, \& {Eta Car HST
  Treasury Team}}]{2003AAS...203.5802C}
{Corcoran}, M.~F., {Hamaguchi}, K., {Henley}, D., {et~al.} 2003, American
  Astronomical Society Meeting Abstracts, 203

\bibitem[{{De Marco} \& {Schmutz}(1999)}]{1999A&A...345..163D}
{De Marco}, O. \& {Schmutz}, W. 1999, \aap, 345, 163

\bibitem[{{De Marco} {et~al.}(2000){De Marco}, {Schmutz}, {Crowther},
  {Hillier}, {Dessart}, {de Koter}, \& {Schweickhardt}}]{2000A&A...358..187D}
{De Marco}, O., {Schmutz}, W., {Crowther}, P.~A., {et~al.} 2000, \aap, 358, 187

\bibitem[{{Dessart} {et~al.}(2000){Dessart}, {Crowther}, {Hillier}, {Willis},
  {Morris}, \& {van der Hucht}}]{2000MNRAS.315..407D}
{Dessart}, L., {Crowther}, P.~A., {Hillier}, D.~J., {et~al.} 2000, \mnras, 315,
  407

\bibitem[{{Hanbury Brown} {et~al.}(1970){Hanbury Brown}, {Davis},
  {Herbison-Evans}, \& {Allen}}]{1970MNRAS.148..103H}
{Hanbury Brown}, R., {Davis}, J., {Herbison-Evans}, D., \& {Allen}, L.~R. 1970,
  \mnras, 148, 103

\bibitem[{{Hanson} {et~al.}(1996){Hanson}, {Conti}, \&
  {Rieke}}]{1996ApJS..107..281H}
{Hanson}, M.~M., {Conti}, P.~S., \& {Rieke}, M.~J. 1996, \apjs, 107, 281

\bibitem[{{Henley} {et~al.}(2005){Henley}, {Stevens}, \&
  {Pittard}}]{2005MNRAS.356.1308H}
{Henley}, D.~B., {Stevens}, I.~R., \& {Pittard}, J.~M. 2005, \mnras, 356, 1308

\bibitem[{{Hillier} \& {Miller}(1998)}]{1998ApJ...496..407H}
{Hillier}, D.~J. \& {Miller}, D.~L. 1998, \apj, 496, 407

\bibitem[{{Malbet} {et~al.}(2005){Malbet}, {Benisty}, {De Wit}, {Kraus},
  {Meilland}, {Millour}, {Tatulli}, {Berger}, {Chesneau}, {Hofmann}, {Isella},
  {Natta}, {Petrov}, {Preibisch}, {Stee}, {Testi}, {Weigelt}, \& {AMBER
  Collaboration}}]{2005-AA-MWC297}
{Malbet}, F., {Benisty}, M., {De Wit}, W.~J., {et~al.} 2005, \aap, -, In Press

\bibitem[{{Martins} {et~al.}(2005){Martins}, {Schaerer}, \&
  {Hillier}}]{2005A&A...436.1049M}
{Martins}, F., {Schaerer}, D., \& {Hillier}, D.~J. 2005, \aap, 436, 1049

\bibitem[{{Mege} {et~al.}(2000){Mege}, {Malbet}, \&
  {Chelli}}]{2000SPIE.4006..299M}
{Mege}, P., {Malbet}, F., \& {Chelli}, A. 2000, in Proc. SPIE Vol. 4006, p.
  299-307, Interferometry in Optical Astronomy, Pierre J. Lena; Andreas
  Quirrenbach; Eds., ed. P.~J. {Lena} \& A.~{Quirrenbach}, 299--307

\bibitem[{{Millour} {et~al.}(2004){Millour}, {Tatulli}, {Chelli}, {Duvert},
  {Zins}, {Acke}, \& {Malbet}}]{2004SPIE.5491.1222M}
{Millour}, F., {Tatulli}, E., {Chelli}, A.~E., {et~al.} 2004, in New Frontiers
  in Stellar Interferometry., ed. W.~A. Traub., Vol. 5491 (SPIE), 1222

\bibitem[{{Millour} {et~al.}(2006){Millour}, {Vannier}, {Petrov}, {Chesneau},
  {Dessart}, \& {Stee}}]{2006-ITHD_FMillour}
{Millour}, F., {Vannier}, M., {Petrov}, R.~G., {et~al.} 2006, in Astronomy with
  High Contrast Imaging III: Instrumentation and data processing, ed.
  C.~{Aime}, A.~{Ferrari}, \& M.~{Carbillet} (EAS Publications Series), in
  press

\bibitem[{{Monnier} {et~al.}(2002){Monnier}, {Greenhill}, {Tuthill}, \&
  {Danchi}}]{2002ASPC..260..331M}
{Monnier}, J.~D., {Greenhill}, L.~J., {Tuthill}, P.~G., \& {Danchi}, W.~C.
  2002, in ASP Conf. Ser. 260: Interacting Winds from Massive Stars, ed.
  A.~F.~J. {Moffat} \& N.~{St-Louis}, 331

\bibitem[{{Perrin}(2003)}]{2003A&A...400.1173P}
{Perrin}, G. 2003, \aap, 400, 1173

\bibitem[{{Petrov} {et~al.}(2003){Petrov}, {Malbet}, {Weigelt}, {Lisi},
  {Puget}, {Antonelli}, {Beckmann}, {Lagarde}, {Lecoarer}, {Robbe-Dubois},
  {Duvert}, {Gennari}, {Chelli}, {Dugue}, {Rousselet-Perraut}, {Vannier}, \&
  {Mourard}}]{2003SPIE.4838..924P}
{Petrov}, R.~G., {Malbet}, F., {Weigelt}, G., {et~al.} 2003, in Interferometry
  for Optical Astronomy II. Edited by Wesley A. Traub . Proceedings of the
  SPIE, Volume 4838, pp. 924-933 (2003)., ed. W.~A. {Traub}, 924--933

\bibitem[{{Pittard} \& {Stevens}(2002)}]{2002A&A...388L..20P}
{Pittard}, J.~M. \& {Stevens}, I.~R. 2002, \aap, 388, L20

\bibitem[{{Pozzo} {et~al.}(2000){Pozzo}, {Jeffries}, {Naylor}, {Totten},
  {Harmer}, \& {Kenyon}}]{2000MNRAS.313L..23P}
{Pozzo}, M., {Jeffries}, R.~D., {Naylor}, T., {et~al.} 2000, \mnras, 313, L23

\bibitem[{{Schaerer} {et~al.}(1997){Schaerer}, {Schmutz}, \&
  {Grenon}}]{1997ApJ...484L.153S}
{Schaerer}, D., {Schmutz}, W., \& {Grenon}, M. 1997, \apjl, 484, L153

\bibitem[{{Schild} {et~al.}(2004){Schild}, {G\" udel}, {Mewe}, {Schmutz},
  {Raassen}, {Audard}, {Dumm}, {van der Hucht}, {Leutenegger}, \&
  {Skinner}}]{2004A&A...422..177S}
{Schild}, H., {G\" udel}, M., {Mewe}, R., {et~al.} 2004, \aap, 422, 177

\bibitem[{{Schmutz} {et~al.}(1997){Schmutz}, {Schweickhardt}, {Stahl}, {Wolf},
  {Dumm}, {Gang}, {Jankovics}, {Kaufer}, {Lehmann}, {Mandel}, {Peitz}, \&
  {Rivinius}}]{1997A&A...328..219S}
{Schmutz}, W., {Schweickhardt}, J., {Stahl}, O., {et~al.} 1997, \aap, 328, 219

\bibitem[{{Skinner} {et~al.}(2001){Skinner}, {G\" udel}, {Schmutz}, \&
  {Stevens}}]{2001ApJ...558L.113S}
{Skinner}, S.~L., {G\" udel}, M., {Schmutz}, W., \& {Stevens}, I.~R. 2001,
  \apjl, 558, L113

\bibitem[{{St.-Louis} {et~al.}(1987){St.-Louis}, {Drissen}, {Moffat},
  {Bastien}, \& {Tapia}}]{1987ApJ...322..870S}
{St.-Louis}, N., {Drissen}, L., {Moffat}, A.~F.~J., {Bastien}, P., \& {Tapia},
  S. 1987, \apj, 322, 870

\bibitem[{{St.-Louis} {et~al.}(1993){St.-Louis}, {Willis}, \&
  {Stevens}}]{1993ApJ...415..298S}
{St.-Louis}, N., {Willis}, A.~J., \& {Stevens}, I.~R. 1993, \apj, 415, 298

\bibitem[{{Stevens} {et~al.}(1996){Stevens}, {Corcoran}, {Willis}, {Skinner},
  {Pollock}, {Nagase}, \& {Koyama}}]{1996MNRAS.283..589S}
{Stevens}, I.~R., {Corcoran}, M.~F., {Willis}, A.~J., {et~al.} 1996, \mnras,
  283, 589

\bibitem[{{Tatulli} {et~al.}(2006){Tatulli}, Millour, Chelli, Duvert, Acke,
  Kraus, Malbet, Petrov, Zins, Antonelli, Beckmann, Bresson, Accardo, Agabi,
  Arezki, Aristidi, Baffa, Behrend, Bl¨ocker, Bonhomme, Busoni, Cassaing,
  Clausse, Connot, Delboulb´e, Driebe, Dugu´e, Feautrier, Ferruzzi, Forveille,
  Fossat, Foy, Fraix-Burnet, Gallardo, Gennari, Glentzlin, Gl¨uck, Giani, Gil,
  Heiden, Heininger, Hofmann, Kamm, Kern, Lagarde, Coarer, Contel, Contel,
  Lisi, Lopez, Magnard, Marconi, Mars, Martinot-Lagarde, Mathias, Monin,
  Mouillet, Mourard, M\`ege, Nussbaum, Ohnaka, Pacini, Perraut, Perrier, Puget,
  Rabbia, Rebattu, Reynaud, Richichi, Robbe-Dubois, Roussel, Sacchettini,
  Salinari, Schertl, Solscheid, Stee, Stefanini, Tallon, Tallon-Bosc, Tasso,
  Testi, Valtier, Vannier, Ventura, \& Weigelt}]{2006_Tatulli}
{Tatulli}, E., Millour, F., Chelli, A., {et~al.} 2006, \aap, -, In Press

\bibitem[{{Van der Hucht}(2002)}]{2002Ap&SS.281..199V}
{Van der Hucht}, K.~A. 2002, \apss, 281, 199

\bibitem[{{Van der Hucht} {et~al.}(1996){Van der Hucht}, {Morris}, {Williams},
  {Setia Gunawan}, {Beintema}, {Boxhoorn}, {de Graauw}, {Heras}, {Kester},
  {Lahuis}, {Leech}, {Roelfsema}, {Salama}, {Valentijn}, \&
  {Vandenbussche}}]{1996A&A...315L.193V}
{Van der Hucht}, K.~A., {Morris}, P.~W., {Williams}, P.~M., {et~al.} 1996,
  \aap, 315, L193

\bibitem[{{Van der Hucht} {et~al.}(1997){Van der Hucht}, {Schrijver},
  {Stenholm}, {Lundstrom}, {Moffat}, {Marchenko}, {Seggewiss}, {Setia Gunawan},
  {Sutantyo}, {van den Heuvel}, {de Cuyper}, \& {Gomez}}]{1997NewA....2..245V}
{Van der Hucht}, K.~A., {Schrijver}, H., {Stenholm}, B., {et~al.} 1997, New
  Astronomy, 2, 245

\bibitem[{{Villar-Sbaffi} {et~al.}(2005){Villar-Sbaffi}, {St-Louis}, {Moffat},
  \& {Piirola}}]{2005ApJ...623.1092V}
{Villar-Sbaffi}, A., {St-Louis}, N., {Moffat}, A.~F.~J., \& {Piirola}, V. 2005,
  \apj, 623, 1092

\bibitem[{{Willis} {et~al.}(1995){Willis}, {Schild}, \&
  {Stevens}}]{1995A&A...298..549W}
{Willis}, A.~J., {Schild}, H., \& {Stevens}, I.~R. 1995, \aap, 298, 549

\end{thebibliography}
\bibliographystyle{aa}

\appendix

\section{Data processing status}\label{section:observingContext}

\begin{figure*}[htbp]
  \centering
  \begin{tabular}{cc}
    \includegraphics[width=0.45\textwidth]{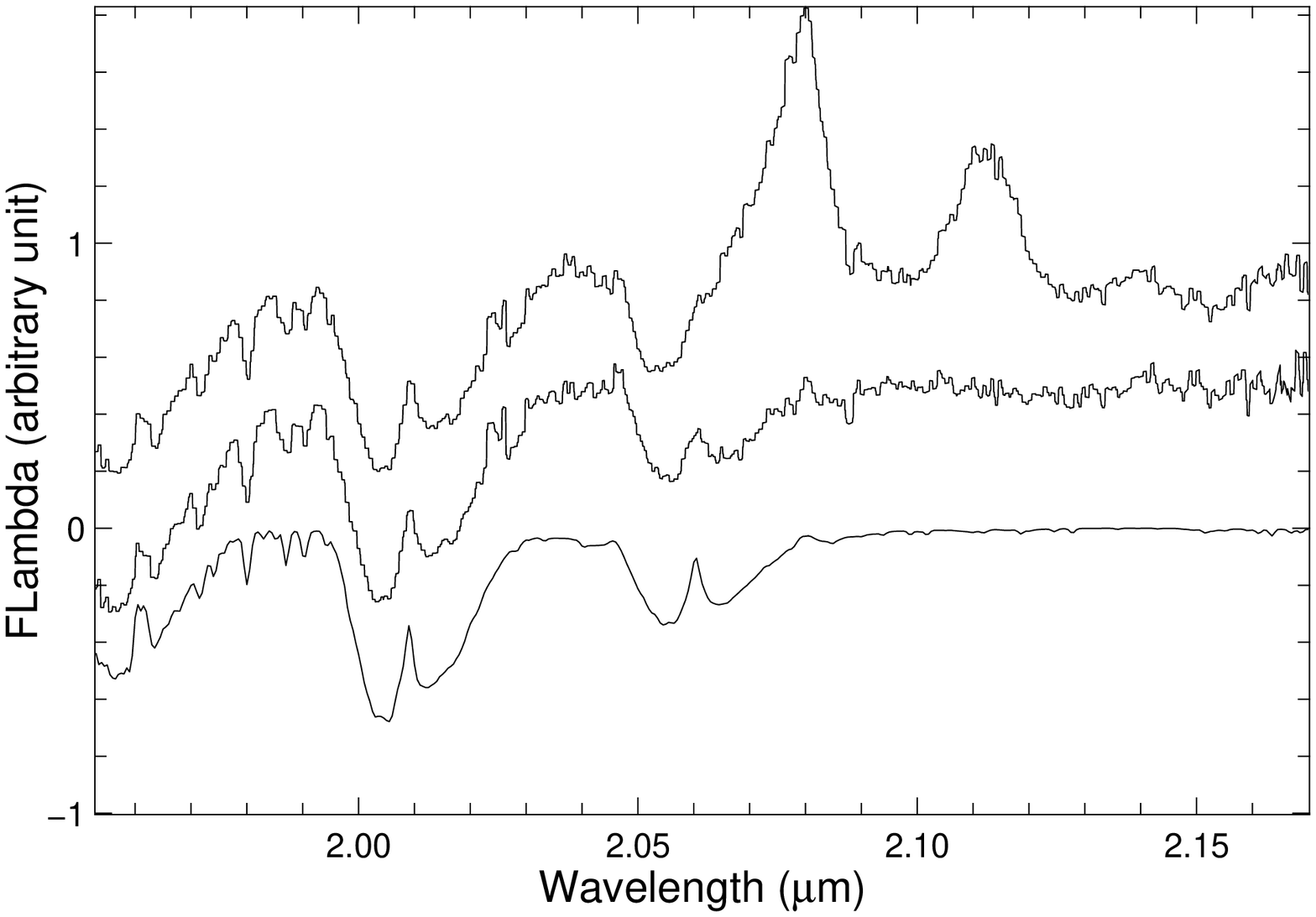}&
    \includegraphics[width=0.45\textwidth]{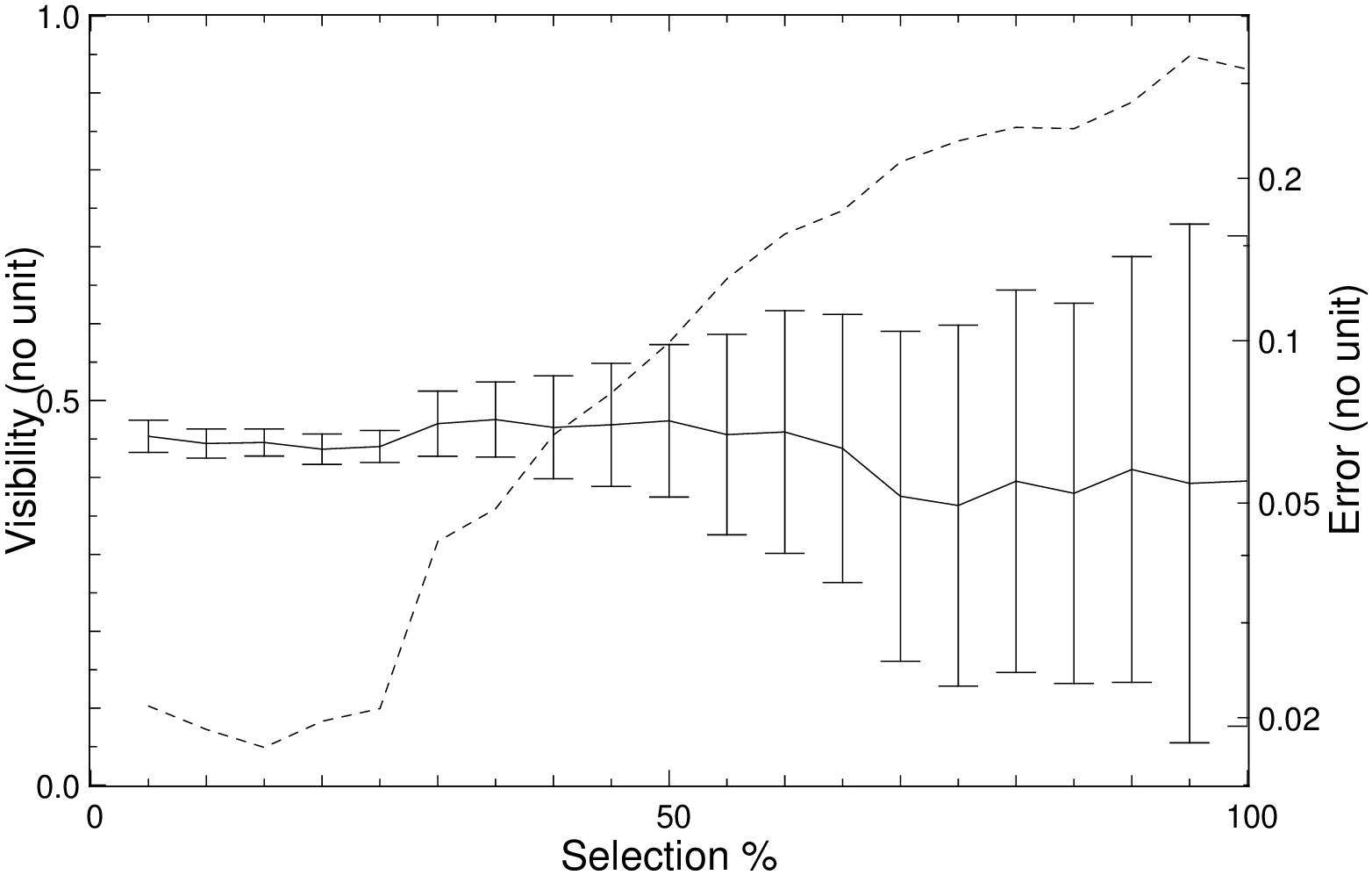}\\
  \end{tabular}
  \caption{
    \footnotesize{
      {\it Left}:
      Calibration of the spectral drift, using the reference star
      spectrum (flat A1{\sc iii} star spectrum). From top to bottom
      are the uncalibrated observed $\gamma^2$~Velorum spectrum, the
      calibrator star spectrum after removal of a Voigt profile in the
      Br$\gamma$ line, and the reference Gemini spectrum. All of them
      show a clear CO$_2$ rovibrationnal feature at 2.01,$\mu$m and 
      2.06\,$\mu$m, and some other water vapor absorption lines, used for the
      absolute wavelength calibration. The graphs have been offset
      for clarity.
      {\it Right}:
      Estimate of the absolute visibilities by applying different
      frame selection thresholds keeping percentage of the observed
      frames (solid line with error bars, left axis). The internal
      dispersion of the visibilities increases as the amount of data
      rejected decreases but there is no obvious bias coming from data
      selection as the absolute visibilities stays constant within the
      error bars. The optimal selection threshold is chosen at the
      position of minimum of the statistical dispersion of the squared
      visibilities (dashed line, right axis): in this data set, this
      level keeps 20\% of the observed frames (i.e. 80\% rejected).
    }
  }
  \label{figure:SpecCalib/select}
\end{figure*}

AMBER follows the standard data flow system implemented at ESO/VLT
During data acquisition, the software records the images of the
spectrally dispersed fringes as well as those of the telescope
beams.

The standard data reduction method developed and optimized for AMBER
is called P2VM for Pixel-To- Visibilities Matrix \citep{2006_Tatulli,
  2004SPIE.5491.1222M}. The P2VM is a linear matrix method which
computed raw visibilities from AMBER data for each spectral
channel. The P2VM is computed after an internal calibration procedure
which is performed every time the instrument configuration
changes. The complex coherent fluxes are given by the product of the
fluxes measured on each pixel of the detector by this P2VM
matrix. We then compute all the useful observables, namely the
visibility, the closure phase, the differential visibility, and the
differential phase. For more details, please refer to
\citet{2006_Tatulli, 2004SPIE.5491.1222M}.





\section{Spectral calibration}\label{section:SpecCalib}

The atmosphere imprints its signature on the observed spectrum of
$\gamma^2$~Velorum, namely through the characteristic CO$_2$
rovibrationnal lines at 2.01\,$\mu$m and 2.06\,$\mu$m. Using a
reference spectrum of the atmospheric transmission and correlation
techniques, we get an absolute and accurate (half a pixel, to be
compared to the 2 pixels sampling of the spectrum) spectral
calibration of the observed spectrum (see Fig.~\ref{figure:SpecCalib/select})


We corrected the spectra for telluric lines by observing the
calibration star approximately at the same airmass and dividing the
two spectra (the same technique as \citet{1996ApJS..107..281H}).

As we observed at medium spectral resolution ($R \sim 1500$), the
numerous narrow spectral features in the A star spectrum are smeared
out, with the exception of $Br\gamma$. The calibrator spectrum is therefore
featureless and allows a good correction of the telluric lines. The
$Br\gamma$ line at 2.165\,$\mu$m is removed from the calibrator spectrum
by fitting a Voigt profile (see the left panel of Fig.~\ref{figure:SpecCalib/select}).


No accurate correction from the telluric spectrum is performed in the area
of the $Br\gamma$ line due to the narrowness of the spectral window
that lack strong telluric lines. The airmass
was 1.2 for $\gamma^2$~Velorum and 1.1 for the calibrator star which
leads to an error on the calibrated spectrum of about 7\% in the parts
where there is strong atmosphere absorption lines (e.g. around
2.00\,$\mu$m). This systematic error is taken into account in the
calibrated spectrum error bars.

AMBER collects the stellar fluxes through optical fibers. Taking into
account the rapidly varying tip/tilt effect on the fiber entrance, the
changes in airmass, seeing conditions, and vibrations conditions
between the star and calibrator, it is not possible to extract a
reliable absolute flux from the observed star. The observed spectrum
is continuum corrected by  means of a spline curve that passes through
designated continuum regions. This yields a totally flat observed
spectrum as in the top panel of Fig.~\ref{figure:Spectrum}.
The error bars of the resulting spectrum takes into account the
detector noise and the photon noise, as well as the air mass mismatch
between the calibrator and the science stars.


\section{Data selection and biases}

The specific conditions of observations described above make the
selection of the data sample difficult. The data reduction technique
provides accurate results when the sub-set of good frames selected for
the  science and calibration object are unbiased. In our case, the
science and  calibration stars have only one magnitude difference and
the average seeing was 0.65$\arcsec$ for $\gamma^2$~Velorum and
0.7\,$\arcsec$ for the calibration star.


To select the data, we compute all the observables and then estimate
the biases introduced on the absolute visibilities by comparing
different frame selection thresholds (see
Fig.~\ref{figure:SpecCalib/select}). The selection criterion is set by a
threshold value of the coherent flux signal-to-noise ratio (SNR) for
individual frames. We obtain a constant value of visibility whatever
the threshold, which suggests that the frame selection  criterion does
not bias the estimated $\gamma^2$~Velorum absolute visibilities in the
range of the estimated error bars.


We based our optimum SNR threshold selection on the minimum of the 
statistical errors computed on the absolute visibilities. This 
optimum threshold keeps 20\% of the total number of frames 
(Fig.~\ref{figure:SpecCalib/select}, left panel).

\section{Observables calibration}
We calibrate the absolute visibilities using the technique described
in \citet{2003A&A...400.1173P}: we interpolate the calibrator
visibilities at the time of the science star observations in order to
correct it from the instrumental and atmospheric transfer
functions. The calibrator visibilities are corrected from the resolved
flux level based on its estimated angular diameter.

%

\begin{equation}
  V^{\rm GV} =  2 \frac {J_1(  \frac{\pi B \theta}{\lambda} ) }{\frac{\pi B \theta}{\lambda}}
  \frac {V^{\rm GV}_{\rm raw}}{V^{\rm cal}}\,,
  \label{eqn_Visibilite}
\end{equation}

where $B$ is the base length, $\theta_{\rm cal}$ the estimated diameter of the
calibration star, $V^{\rm GV}_{\rm raw}$ being the raw visibilities
of $\gamma^2$~Velorum, and $V^{\rm cal}$ the visibility of the
calibration star interpolated to the observation time.

For the closure phase, we only correct the object closure phase from any 
instrumental-based signal by subtracting the calibrator closure phase
and the object closure phase,

\begin{equation}
  \psi^{\rm GV}_{\rm diff} = \psi^{\rm GV}_{\rm diff\,raw} - \psi^{\rm
    cal}_{\rm diff\,raw}\,.
  \label{eqn_Closure}
\end{equation}

The differential visibilities and differential phase are computed as
explained in \citet{2006-ITHD_FMillour}, and should need no
calibration, at least in theory. However, we noticed that they are
affected by instrumental effects such as polarization mismatch between
the fibers outputs, leading to differential phase effects up to
0.05\,rad peak to peak.

For the differential visibility, the calibration is performed by
dividing the star by the calibrator differential visibilities:

\begin{equation}
  V^{\rm GV}_{\rm diff} = \frac {V^{\rm GV}_{\rm diff\,raw}}{V^{\rm
      cal}_{\rm diff\,raw}}
  \label{eqn_VisibiliteDiff}
\end{equation}

For the differential phase, the calibration is performed by
substracting the calibrator to the star differential phase:

\begin{equation}
  \phi^{\rm GV}_{\rm diff} = \phi^{\rm GV}_{\rm diff\,raw} - \phi^{\rm
    cal}_{\rm diff\,raw}
  \label{eqn_PhaseDiff}
\end{equation}



\section{Error Estimates}   {

  The statistical errors are estimated using the dispersion of the
  individual-frame observables, assuming a Gaussian distribution of
  differential visibilities, differential phases and closure phase.

  The limitations of the VLTI + AMBER instrument described in
  Appendix.~\ref{section:observingContext} (vibrations, large amount of
  time between the star and calibrator) and the level of dispersion of
  the absolute visibilities observed with other calibrators (in
  different spectral band) led us to increase the estimated error bars
  beyond the natural dispersion of the absolute visibilities. We found
  that the errors between calibrators are typically 5\% above the
  internal dispersion, and, thus, this error has been added to the
  error budget. Hence, the absolute visibility error bars contain two
  contributions: the statistical dispersion of the measured spectrally
  dispersed absolute visibilities and the bias error of the mean
  visibilities in a spectral window, estimated to be 5\%.

\end{document}